\documentclass[sigconf]{acmart}
\AtBeginDocument{%
  }

\copyrightyear{2026}
\acmYear{2026}
\setcopyright{cc}
\setcctype{by}
\acmConference[UIST '26]{The 39th Annual ACM Symposium on User Interface Software and Technology}{November 02--05, 2026}{Detroit, MI, USA}
\acmBooktitle{The 39th Annual ACM Symposium on User Interface Software and Technology (UIST '26), November 02--05, 2026, Detroit, MI, USA}
\acmDOI{10.1145/3830398.3830638}
\acmISBN{979-8-4007-2856-3/2026/11}

\usepackage{xspace}
\usepackage{subcaption}
\usepackage{fancyvrb}
\usepackage{fvextra}
\usepackage{float}
\usepackage{balance}

\newcommand{\systemName}{\textsc{Graphionale}\xspace}

\newcommand{\rvs}[1]{\textcolor{black}{#1}}
\definecolor{mypurple}{HTML}{76448A}
\definecolor{mygreen}{HTML}{0E6655}

\begin{document}

\title{\systemName: How Graph Visualizations of LLM Rationales Affect Human Decision Making}


\author{Xinru Wang}
\authornote{Both authors contributed equally to this research.}
\orcid{0000-0002-0213-6425}
\affiliation{%
  \institution{Singapore-MIT Alliance for Research and Technology}
  \country{Singapore}}
\email{xinru.wang@smart.mit.edu}

\author{Zhexuan Ma}
\authornotemark[1]
\orcid{0009-0007-7118-9778}
\affiliation{%
  \institution{National University of Singapore}
  \country{Singapore}}
\email{e1499160@u.nus.edu}

\author{Ming Yin}
\orcid{0000-0002-7364-139X}
\affiliation{%
  \institution{Purdue University}
  \city{West Lafayette}
  \state{Indiana}
  \country{USA}}
\email{mingyin@purdue.edu}

\author{Shuai Ma}
\authornote{Corresponding author.}
\orcid{0000-0002-7658-292X}
\affiliation{%
  \institution{
  Institute of Software, 
  Chinese Academy of Sciences}
  \city{Beijing}
  \country{China}}
\email{mashuai@iscas.ac.cn}

\author{Thomas W Malone}
\orcid{0000-0002-7005-1482}
\affiliation{%
  \institution{
  Massachusetts Institute of Technology}
  \city{Cambridge}
  \state{Massachusetts}
  \country{USA}}
\email{malone@mit.edu}

\renewcommand{\shortauthors}{Wang et al.}

\begin{abstract}
Large Language Models (LLMs) are increasingly equipped with augmented reasoning capabilities to generate rationales that support human decision-making. Yet these text-dense rationales often impose substantial cognitive burdens. 
Building on a formative co-design study that identified user preferences for non-linear reasoning representations, \rvs{we developed \systemName\ as a testbed for empirically studying argument-map-style rationale visualization. This system} transforms linear LLM rationales into interactive, multi-level graphs.
\rvs{It} explicitly structures logical relationships (e.g., conclusions, premises, support, and objections), while further extracting entities and relations within each statement to construct condensed node-link representations.
We conduct a large-scale online user study ($N=204$) to examine when graphical rationales are more effective than textual ones, across varying task modality (verbal vs.\ visual reasoning), rationale format (textual vs.\ graphical), and question difficulty (easy vs.\ hard). Our results show that graphical rationales \rvs{do not help uniformly: they} improve trust calibration for verbal reasoning yet feel more cognitively demanding and less satisfying; for visual reasoning, they impair calibration yet feel more engaging and helpful. \rvs{In each modality, the format that better supports calibrated decisions is not the one users prefer,} highlighting that matching rationale format to task modality is key to effective AI explanation design.
Our findings \rvs{contribute empirical design knowledge about when and how graphical rationales support human decision making, and inform the} next-generation reasoning-aware AI interfaces.
\end{abstract}


\begin{teaserfigure}
\centering
  \includegraphics[width=0.97\textwidth]{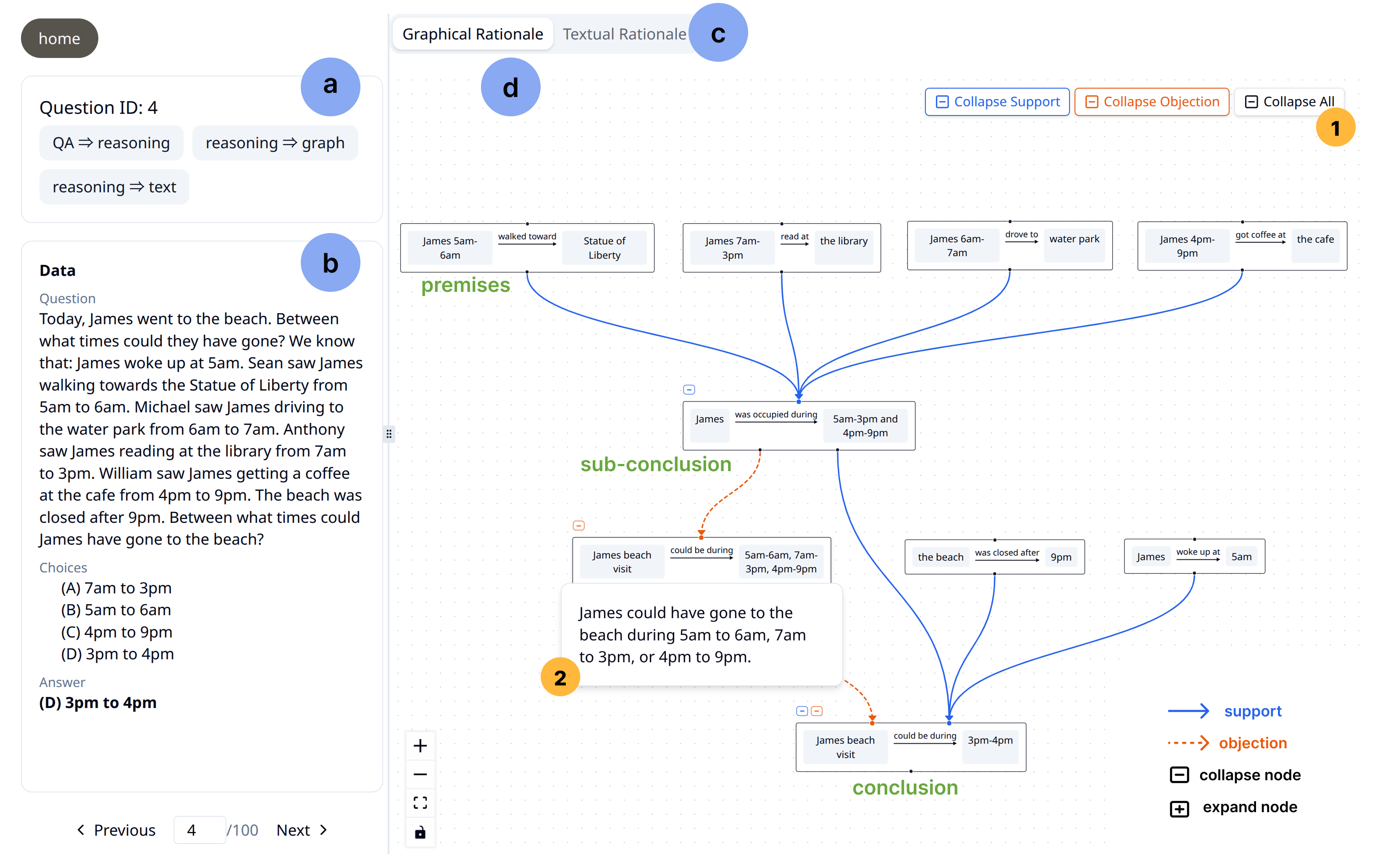}
  \caption{
    \systemName, \rvs{a co-design-informed prototype that} transforms LLM rationales into structured argument graphs with compressed node representations. 
    The interface supports textual and graphical rationale generation (a), task question display (b), switching between views (c), and an interactive graph panel (d). Users can selectively expand or collapse reasoning branches through global controls (1) and inspect detailed explanations by hovering over nodes (2).}
  \Description{Screenshot of the \systemName interface showing a BBH Temporal Sequences task. The left panel (a) displays the task question, answer choices (A--D), and pipeline controls for generating the reasoning structure and rationales. The right panel shows the graphical rationale view (c), with global collapse controls for support and objection branches in the top-right corner (1). The graph contains four premise nodes at the top encoding time-stamped observations about James's activities, connected via blue solid support edges to an intermediate sub-conclusion node (James was occupied during 5am--3pm and 4pm--9pm), which further connects to an objection node (James beach visit could be during 5am--6am, 7am--3pm, or 4pm--9pm) via an orange dashed objection edge. Additional premises (the beach was closed after 9pm, James woke up at 5am) lead to the final conclusion node at the bottom (James beach visit could be during 3pm--4pm). A legend on the right explains the edge styles (support, objection) and node controls (collapse, expand). Hovering over a node (2) reveals the full-text explanation of that reasoning step.}
  \label{fig:teaser}
\end{teaserfigure}


\maketitle

\section{Introduction}
Recent advances in Large Language Models (LLMs) have significantly improved their reasoning capabilities.
With these developments, LLMs are increasingly used as conversational assistants for complex, knowledge-intensive questions. To improve transparency and support human decision making, LLMs provide detailed rationales describing how an answer is derived~\cite{wei_chain--thought_2022,yao_tree_2023,besta_graph_2024,yao_react_2022,muennighoff_s1_2025}. However, these rationales are traditionally presented as long blocks of text. Even when organized into sections, text remains inherently linear and cannot intuitively represent relational structures among reasoning steps, such as branching, aggregation, or elimination. As a result, users must parse sequential sentences to reconstruct the underlying logic, making it difficult to quickly grasp the overall reasoning or evaluate the model's conclusions~\cite{jiang_graphologue_2023,pang_interactive_2025}.

In contrast, humans often externalize complex explanations into diagrams to support understanding. Students draw concept maps to visualize relationships between ideas~\cite{novak_theory_2008,ruiz-primo_problems_1996}. In deliberation settings, discussions are represented as argument maps to help people navigate competing viewpoints~\cite{conklin_gibis_1988,klein_how_2007}.
Inspired by these practices, representing LLM rationales graphically---by making logical relationships explicit and visually structured---may better align with human reasoning processes and support users in understanding and critically evaluating model outputs.
\rvs{However, graphical rationales may also make flawed or unfaithful LLM reasoning appear more credible than it is~\cite{mishra_characterizing_2024}. Whether they help users identify such problems or instead inflate trust remains an empirical question.}

Several recent systems have begun exploring graphical interfaces for LLM outputs, including tools for prompt engineering~\cite{arawjo_chainforge_2024}, interactive diagrams of LLM responses~\cite{jiang_graphologue_2023,suh_sensecape_2023}, and reasoning trace visualizations~\cite{pang_interactive_2025,li_reasongraph_2025}. However, the design principles for graphical LLM rationales remain underexplored, and little empirical evidence exists on whether, when, and how such representations benefit users.

To address this gap, we investigate two research questions:

\begin{itemize}
    \item \textbf{RQ1:} How should graphical representations of LLM rationales be designed to align with user needs?
    \item \textbf{RQ2:} Compared with textual rationales, how do graphical LLM rationales influence users' interaction experience and decision-making outcomes?
\end{itemize}

To answer RQ1, we conducted a formative co-design study in which participants interacted with a low-fidelity interface. Through sketching and discussion, we examined how users expect LLM rationales to be graphically represented. Guided by argument mapping theory~\cite{wikipedia_argument_2026} and insights from this study, we developed \systemName\footnote{\url{https://graphionale.vercel.app/}}, \rvs{an LLM-specific instantiation of argument-map-style rationale visualization} that generates glanceable graphical representations of LLM rationales by (1) structuring reasoning as \rvs{argument maps} and (2) compressing node-level text to highlight essential information~\cite{wang_less_2025,jiang_graphologue_2023}. \rvs{\systemName serves as a co-design-informed prototype for the empirical comparison that follows.}

To answer RQ2, we conducted a controlled online user study ($N=204$) comparing graphical and textual LLM rationales across different decision-making contexts. We varied task modality (verbal vs.\ visual reasoning) and question difficulty (easy vs.\ hard) to examine how these factors influence the effectiveness of graphical rationales.
Our results show that graphical rationales \rvs{do not help uniformly: contrary to intuitive expectations, they} improve trust calibration for verbal reasoning yet are rated as more cognitively demanding and less satisfying; for visual reasoning, graphical rationales impair calibration yet are rated as more engaging and helpful. \rvs{In each modality, the format that supports better-calibrated decisions is thus not the one users prefer.} This objective--subjective dissociation suggests that perceived fluency does not reliably predict calibration quality. Furthermore, the calibration benefit in verbal reasoning is amplified on harder questions and among users higher in analytical thinking.

In summary, this paper makes the following contributions:
\begin{itemize}
    \item A formative co-design study grounded in argument mapping that identifies key design principles for presenting LLM rationales graphically.
    \item \systemName, \rvs{a co-design-informed prototype that instantiates argument-map-style visualization for LLM rationales} by combining structured argument mapping with text condensation\rvs{, serving as the vehicle for our empirical comparison}.
    \item A large-sample controlled user study that systematically evaluates, compared with traditional textual rationales, how graphical LLM rationales influence user perception and behavior across task modalities, \rvs{contributing empirical design knowledge about when and how graphical rationales support human decision making}.
\end{itemize}

\section{Related Work}
\subsection{Graph-driven Explainable AI and Reasoning in LLMs}

Graph structures---particularly Knowledge Graphs (KGs)---have long been used to provide transparency in AI systems. Research has shown that inferring preference paths over user--item--entity graphs can generate human-readable reasoning chains~\cite{geng_path_2022,song_ekar_2022}, and KGs have been applied for causal and counterfactual reasoning~\cite{jaimini_causalkg_2022,rajabi_knowledge-graph-based_2024}.

In the context of LLMs, the focus has evolved toward multi-hop explanation graphs. Datasets such as WorldTree~\cite{jansen_worldtree_2018} formalize rationales as lexically connected facts. To augment LLM reasoning, researchers have moved beyond linear Chain-of-Thought prompting~\cite{wei_chain--thought_2022} toward structured approaches such as Tree-of-Thought~\cite{yao_tree_2023} and Graph-of-Thought~\cite{besta_graph_2024}, which allow models to explore multiple reasoning branches. Recent test-time scaling methods~\cite{muennighoff_s1_2025} further enhance reasoning performance.

Despite these advances, a critical gap persists at the human--AI \textit{interface}: rationales are typically presented as dense text, which obscures their multi-hop structure and limits users' ability to efficiently inspect and evaluate the reasoning process.

\subsection{Graph Structures in Human Mental Models}

Human cognition is inherently graph-structured. Research suggests that humans represent knowledge as associative networks in which sensory, spatial, and semantic nodes are multiply connected. The ability to infer and navigate these relationships is fundamental to forward planning and logical reasoning~\cite{rmus_humans_2022}.

Humans naturally organize semantic knowledge using graph-like structures~\cite{collins_retrieval_1969}. Representations such as mind maps and concept maps leverage perceptual strengths to manage cognitive load, for example by exploiting Gestalt principles such as proximity and continuity~\cite{parsons_understanding_2022}. Visual working memory is typically limited to roughly 4–7 chunks~\cite{ware_information_2019}, motivating design strategies that prioritize task-relevant nodes while compacting less relevant ones. Effective conceptual mapping also relies on hierarchical layering to structure complex information~\cite{novak_theory_2008,ruiz-primo_problems_1996}.

If knowledge provides the cognitive space, reasoning can be understood as tracing paths through it. Deductive, inductive, and abductive reasoning correspond to different ways of selecting and chaining relations. Argument maps provide a structural representation of these inference processes by explicitly linking premises, evidence, and conclusions~\cite{wikipedia_argument_2026}. By organizing arguments into explicit structures, they help clarify complex reasoning, identify logical weaknesses, and support critical evaluation and collaborative discussion.

Early systems such as gIBIS~\cite{conklin_gibis_1988} and platforms like the Deliberatorium~\cite{klein_how_2007} externalize reasoning as navigable graphs, enabling users to synthesize complex information more effectively than through text-based discourse alone and supporting large-scale deliberation and collective intelligence.

\subsection{Design and Evaluation of Graphical XAI and LLM Interfaces}

Researchers have explored graph-based visualizations to support AI explainability in ways that align with human mental models, typically representing explanations as structured relationships among concepts or decision provenance. For example, ConceptExplainer~\cite{huang_conceptexplainer_2023} uses concept graphs to visualize relationships between learned concepts and model predictions. Other work~\cite{jaigirdar_what_2020, madanagopal_analytic_2019} leverages provenance graphs to capture security-aware meta-information for explaining AI systems.

Prior work has also examined how such visualizations affect human understanding. \citet{delarue_evaluating_2024} found that graph-based explanations were perceived as more usable and intuitive than feature-importance explanations in recommender systems, although textual explanations sometimes led to higher objective understanding. This highlights the importance of carefully designing graph representations. \citet{montagna_graph-based_2023} further emphasize the need for rigorous, metric-driven evaluation of graph-based explanation quality (e.g.,~\cite{wang_watch_2023,wang_are_2021,wang_effects_2022,wang_effects_2023,ma_towards_2025,vasconcelos_explanations_2023,gajos_people_2022}), rather than relying solely on qualitative case studies.

Looking at graph representations of LLM-generated text, one line of work focuses on extracting causal structures. \citet{yang_survey_2022} review methods for identifying cause–effect relationships, while recent work explores using LLMs to discover causal structures~\cite{vashishtha_causal_2024} and generate concept maps~\cite{perin_text_2023}. Systems such as PaperTrail~\cite{martin-boyle_papertrail_2026} extract claim–evidence structures from research papers and LLM responses, representing them as provenance graphs for scholarly question answering.

Another line of research explores graphical interfaces for interacting with LLM outputs. Graphologue~\cite{jiang_graphologue_2023} converts responses into interactive diagrams for non-linear exploration, while Sensecape~\cite{suh_sensecape_2023} supports multilevel organization for navigating complex topics. Tools such as ChainForge~\cite{arawjo_chainforge_2024} provide visual environments for prompt engineering and hypothesis testing. These systems primarily facilitate LLM output exploration and content organization.

In contrast, visualizing LLM rationales requires representing implicit reasoning structures---such as dependencies among steps and how conclusions are derived---in a way that is explicit and interpretable. Only recently have efforts begun to address this. HaLLMark~\cite{hoque_hallmark_2024} visualizes authorship provenance, while \citet{wang_less_2025} improve glanceability by structuring text into taxonomy components without modeling reasoning as graphs. ReasonGraph~\cite{li_reasongraph_2025} and HIPPO~\cite{pang_interactive_2025} visualize reasoning traces as graphs or trees with interactive features.
However, these systems lack clear design principles for making graphical rationales glanceable and interpretable, and their evaluations are small-scale usability studies that rarely consider factors such as user reliance on AI or decision-making task characteristics.

\textbf{Research Gap.}
While prior work demonstrates the potential of graph representations for explaining AI systems and exploring LLM outputs, there is limited empirical evidence on how graphical representations of LLM \textit{rationales} can be effectively designed, and how they affect human understanding, decision-making, and human--AI collaboration.

\section{Formative Study: A Co-Design Exploration}
\subsection{Participants and Study Procedure}

To explore what forms of graphical rationales users prefer\rvs{, and to inform the design of a strong argument-map-style representation for our empirical comparison}, we conducted a formative study with co-design activities. We recruited 10 participants from our institution (5 female, 5 male; mean age 30), including 3 graduate students, 6 research scientists, and 1 software engineer. All were familiar with LLM applications and used them daily.

We built a low-fidelity interface that allowed participants to interact with both textual and graphical LLM rationales. We selected XplainLLM~\cite{chen_xplainllm_2024}, in which each instance includes explanatory text describing LLM reasoning on CommonsenseQA~\cite{talmor_commonsenseqa_2019}. Using GPT-5.2 with a prompt adapted from \cite{harrell_creating_2010} (see Appendix~\ref{ap:formative_prompt}), we converted each explanation into a graphical representation where nodes represent statements (premises, intermediate conclusions, final conclusion) and directed edges represent inference relationships. A screenshot of the interface is provided in Appendix~\ref{ap:formative_screenshot}.

During the study, participants completed two randomly selected CommonsenseQA questions. They first reviewed the textual rationale and described any difficulties, then switched to the graphical rationale and compared the two formats. Participants proposed improvements and sketched how they would like the rationale represented graphically. 
Each session lasted around 25 minutes (\$6 compensation). All sessions were recorded for subsequent analysis.

\subsection{Analysis and Results}

Recorded sessions were transcribed and analyzed using thematic analysis~\cite{braun_using_2006}. Two authors iteratively discussed and refined a shared codebook. The analysis revealed four themes describing how users believe graphical representations of LLM rationales should be improved.

\paragraph{\textbf{Theme 1: Glanceable nodes require aggressive text compression.}}
\textcolor{mypurple}{\textbf{Challenge (C1).}}
Participants reported that many nodes felt like \textit{``reading the text again,''} as sentence-level content prevented quick scanning. As P1 noted, \textit{``If this block of text could be split into segments and listed as items, it would be easier.''} When diagrams became overloaded, participants preferred falling back to structured text. As P10 commented, \textit{``It didn't really help me solve the problem. I still had to read a lot of text.''} Participants suggested replacing full sentences with keywords or short phrases and adjusting node granularity.

\textcolor{mygreen}{\textbf{Design Goal (DG1).}}
Enable fast scanning through compressed keyword-based nodes, while retaining structured text as a fallback when diagrams do not improve clarity.

\paragraph{\textbf{Theme 2: Explanation maps should align with human decision logic.}}
\textcolor{mypurple}{\textbf{Challenge (C2).}}
Participants struggled when the map's reasoning direction conflicted with common reading habits. They expected explanations to support elimination and comparison across answer options. As P2 described, \textit{``I would first eliminate obviously wrong options, then compare the remaining ones.''} Some rationales contained irrelevant terms not present in the answer options. P8 reacted: \textit{``This looks like a hallucination\ldots these things are not even in the answer options.''} Participants proposed restructuring explanations around a decision template: criteria $\rightarrow$ comparison $\rightarrow$ elimination $\rightarrow$ summary (example sketches are provided in Appendix~\ref{ap:formative_sketch}), and adapting depth to task difficulty.

\textcolor{mygreen}{\textbf{Design Goal (DG2).}}
Make explanation maps function as decision-support artifacts that mirror human reasoning and option elimination, without exposing unnecessary model internals.

\paragraph{\textbf{Theme 3: Visual semantics are essential for fast interpretation.}}
\textcolor{mypurple}{\textbf{Challenge (C3).}}
Participants had difficulty interpreting relationships when visual semantics were unclear. P2 noted, \textit{``This graph is messy\ldots the lines are confusing, and I can't tell what the relationships are.''} Participants wanted to quickly distinguish supporting and rejecting evidence and understand reasoning strength without reading every node. They proposed adding colors, icons, or symbols and clarifying edge semantics with labels and grouping structures.

\textcolor{mygreen}{\textbf{Design Goal (DG3).}}
Enable fast relational scanning so that support, rejection, grouping, and reasoning strength are visually interpretable at a glance.

\paragraph{\textbf{Theme 4: Complexity must be bounded through constraints and progressive disclosure.}}
\textcolor{mypurple}{\textbf{Challenge (C4).}}
Participants found diagrams overwhelming when too many nodes were displayed. P4 reacted: \textit{``This is way too much\ldots how did it generate so many nodes?''} Navigation costs were high for large diagrams. P3 noted, \textit{``It's so long that I have to keep dragging around, which is exhausting.''} Participants suggested merging redundant nodes, imposing limits on depth, and using progressive disclosure so that details appear only on demand. P1 proposed keeping the structure within \textit{``about three to four layers, with three to five nodes per layer.''}

\textcolor{mygreen}{\textbf{Design Goal (DG4).}}
Keep explanations bounded, navigable, and user-controllable so that diagrams remain glanceable rather than overwhelming.

\section{\systemName\ Interface}
\subsection{System Walkthrough}

We instantiate these design goals in \systemName, as illustrated in the following scenario and in \autoref{fig:teaser}.

Alex is solving a logic puzzle with the help of a chatbot. The task involves determining when a person could have gone to the park based on a series of time-stamped activities (e.g., walking toward the Statue of Liberty from 5am to 6am, driving to the water park from 6am to 7am). In a traditional interface, the rationale appears as multiple paragraphs that Alex must read through and mentally reconstruct.

With \systemName, the rationale is presented as a graphical structure that explicitly organizes statements and their relationships. The graph aligns with human decision logic: constraints are first aggregated, then used to eliminate candidate time ranges, leading to a single valid interval (\textcolor{mygreen}{\textbf{DG2}}). The graph adapts its complexity: high-level constraints are shown first, while details can be expanded on demand (\textcolor{mygreen}{\textbf{DG4}}).

Each node represents a compressed, keyword-based statement for quick scanning (\textcolor{mygreen}{\textbf{DG1}}). Within each node, the statement is structured using subject--predicate--object triplets, e.g., ``5am--6am'' $\xrightarrow{\text{walking toward}}$ ``Statue of Liberty''. Edges visually encode relationships between statements such as support and objection (\textcolor{mygreen}{\textbf{DG3}}).


\subsection{Graphical Rationale Generation}

Given a reasoning question and an answer, we decompose graphical rationale generation into two steps: (1) constructing an argument map, and (2) compressing statement text within each node.

\subsubsection{Argument Map Construction}

Our formative study suggests that people tend to organize premises into intermediate sub-conclusions, progressively derive a final conclusion, and eliminate conflicting alternatives. This reasoning pattern closely matches argument maps~\cite{wikipedia_argument_2026}, which have long been used in education, deliberation, and critical thinking.

Therefore, we adopt argument map principles to guide LLMs in generating structured graphical rationales (\textcolor{mygreen}{\textbf{DG2}}). Specifically, we represent premises, sub-conclusions, and the final conclusion as nodes, and logical relationships as edges, including support (reasons in favor) and objection (reasons against).
The depth of sub-conclusions scales with reasoning complexity (\textcolor{mygreen}{\textbf{DG2}}). Edge color and line style distinguish support and objection relations (\textcolor{mygreen}{\textbf{DG3}}). We constrain the number of nodes and incorporate progressive disclosure for users to expand the graph to reveal deeper reasoning layers or inspect supporting and opposing arguments on demand (\textcolor{mygreen}{\textbf{DG4}}).

\subsubsection{Statement Text Compression}

To further improve glanceability, we compress the textual content within each node. Specifically, we distill each statement into an entity--relation--entity representation (\textcolor{mygreen}{\textbf{DG1}}). The compressed representation captures essential semantic components while reducing verbosity, and is presented in a structured node-link format within each node (\textcolor{mygreen}{\textbf{DG3}}). We impose length constraints for conciseness. As a fallback, users can view the full text on demand. Prompts for argument map construction and statement compression are provided in Appendix~\ref{ap:system_prompt}.

\subsection{Implementation}

\systemName is a full stack web application implemented with Next.js. 
The front-end renders questions and LLM rationales, while the backend manages task data retrieval, generation of graphical rationales, derivation of textual and graphical rationales, and persistence of task data and generated rationales in PostgreSQL.

Rationale generation uses GPT-5.4 with high reasoning effort. The back-end first constructs a structured argument map, then derives both textual and graphical rationales from it. This keeps the two views aligned while supporting different presentation forms.

 The graphical view is built on XYflow (React Flow)~\cite{team_react_2026}, which renders graphical rationales as interactive diagrams with zooming and viewport navigation. We implemented controls for selectively expanding and collapsing support and objection branches, either for individual nodes or globally. By default, each node displays a compact subject--predicate--object triplet when suitable or a short label otherwise. Hovering over a node reveals the full text explanation. 
 An overview of the system architecture is provided in Appendix~\ref{ap:system_architecture}.

\section{Technical Evaluation}

To assess the quality of LLM-generated graphical rationales, we conducted a technical evaluation focusing on the correctness of both graph structure and reasoning content.

\subsection{Setup}

\subsubsection{Dataset Selection}
\label{sec:dataset}

We varied task modality (verbal vs.\ visual reasoning) and question difficulty to improve generalizability, as further explained in Section~\ref{sec:condition}.

For verbal reasoning, we selected BIG-Bench Hard (BBH)~\cite{suzgun_challenging_2023}, a suite of 23 challenging tasks~\cite{authors_beyond_2023} that requires multi-step reasoning. The benchmark reports average human-rater performance, allowing us to select tasks with different difficulty levels. We focused on tasks involving verbal reasoning that remain accessible to lay participants: one easy task with high human performance (``Temporal Sequences'') and two hard tasks with low human performance (``Logical Deduction (7 Objects)'' and ``Tracking Shuffled Objects (7 Objects)''). 

For visual reasoning, we selected I-RAVEN~\cite{hu_stratified_2021,yu_cwhyi-raven_2025}, which improves upon RAVEN~\cite{zhang_raven_2019} for fairer evaluation on Raven's Progressive Matrices. Among its seven figure configurations, we selected ``L-R'' (Left-Right) as the easy task, and ``3$\times$3Grid'' and ``Out-InGrid'' as the hard tasks, since configurations with more components require more complex reasoning. Because I-RAVEN is visual, we prompted the LLM to generate step-by-step graphical logic patterns in SVG-based representations. 
Example questions of BBH and I-RAVEN 
are provided in Appendix~\ref{ap:technical_tasks}.

For each selected question category, we randomly sampled 30 questions. For 20 questions, we generated graphical LLM rationales for the correct answer; for the remaining 10, we generated rationales for a randomly selected incorrect answer.

\subsubsection{Evaluation Criteria}

Two coders evaluated the generated graphical rationales on three aspects:

\begin{itemize}

\item \textbf{Graph generation quality:} Whether generated graphs followed structural constraints~\cite{parsons_understanding_2022,novak_theory_2008,ruiz-primo_problems_1996,li_reasongraph_2025}, including depth scaling with difficulty, node count limits, text length per node, and consistency between compressed nodes and full textual rationales.

\item \textbf{Semantic reasoning quality:} Whether reasoning paths (supporting or rejecting relationships) were logically correct and coherent, and whether node content was valid and interpretable~\cite{li_reasongraph_2025}.

\item \textbf{Functional correctness for error detection:} Whether graphical rationales could help users detect reasoning errors~\cite{hoffman_metrics_2019}---correct answers should produce reasonable rationales, while incorrect answers should reflect flawed or inconsistent reasoning structures.

\end{itemize}

\subsection{Findings}
\label{sec:tech_evaluation_finding}

Coders recorded issues or sub-optimal patterns in generated rationales. Based on these observations, we conducted an iterative prompt refinement process. All findings below are based on the final refined generation. \rvs{Inter-rater reliability was high (Cohen's $\kappa = 0.94$)}.

\textbf{Graph generation quality was high.} Across 180 rationales (30 questions $\times$ 6 task categories), only 5 structural issues (2.8\%) were identified, all involving excessive edges in BBH Logical Deduction questions. No violations were observed in node scaling, node count, text length, or consistency between compressed nodes and full text.

\textbf{Correct-answer rationales had two recurring sub-optimal patterns.} (1) \textit{Redundant reasoning steps} (15/120, 13\%): unnecessary intermediate nodes that added clutter, particularly in I-RAVEN tasks. (2) \textit{Unverifiable reasoning} (10/120, 8\%): nodes containing claims about subtle visual attributes that could not be reliably verified from the image.

\textbf{Three error patterns characterized incorrect-answer rationales.} (1) \textit{Missing evidence} (10/60, 17\%): reasoning chains omitting critical constraints. (2) \textit{Problematic evidence} (26/60, 43\%): hallucinated or factually incorrect content, especially in I-RAVEN. (3) \textit{Incorrect (sub-)conclusions} (30/60, 50\%): valid evidence with incorrect inferences, most prominent in BBH. All three patterns were almost entirely absent in correct-answer rationales, confirming their diagnostic value as indicators of LLM reasoning failure.

\section{User Study Design}
\begin{figure*}[t]
  \centering
  \begin{subfigure}[b]{0.48\textwidth}
    \includegraphics[width=\linewidth]{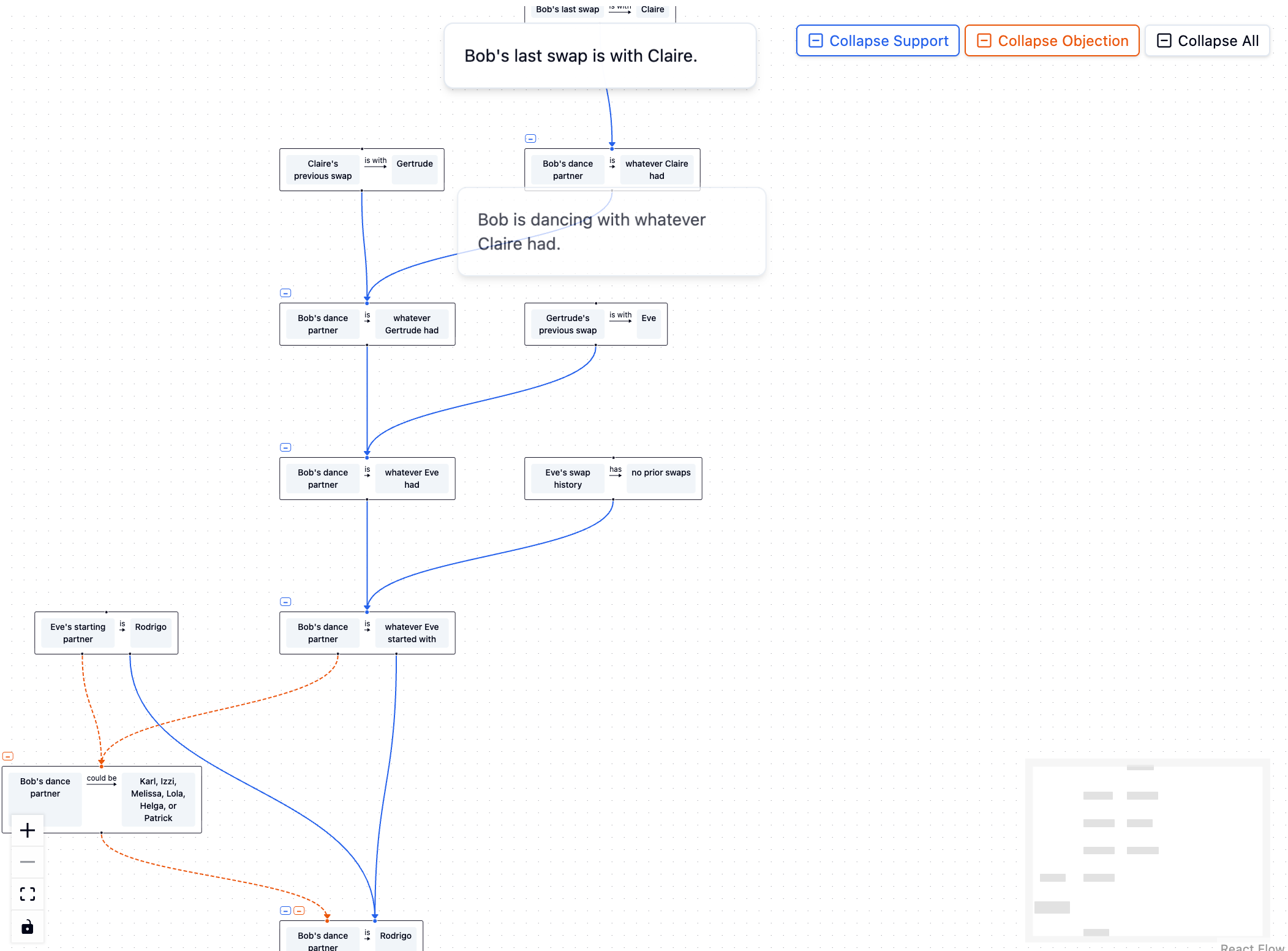}
    \caption{BBH $\times$ Graph}
    \label{fig:screenshot_bbh_graph}
  \end{subfigure}
  \begin{subfigure}[b]{0.48\textwidth}
    \includegraphics[width=\linewidth]{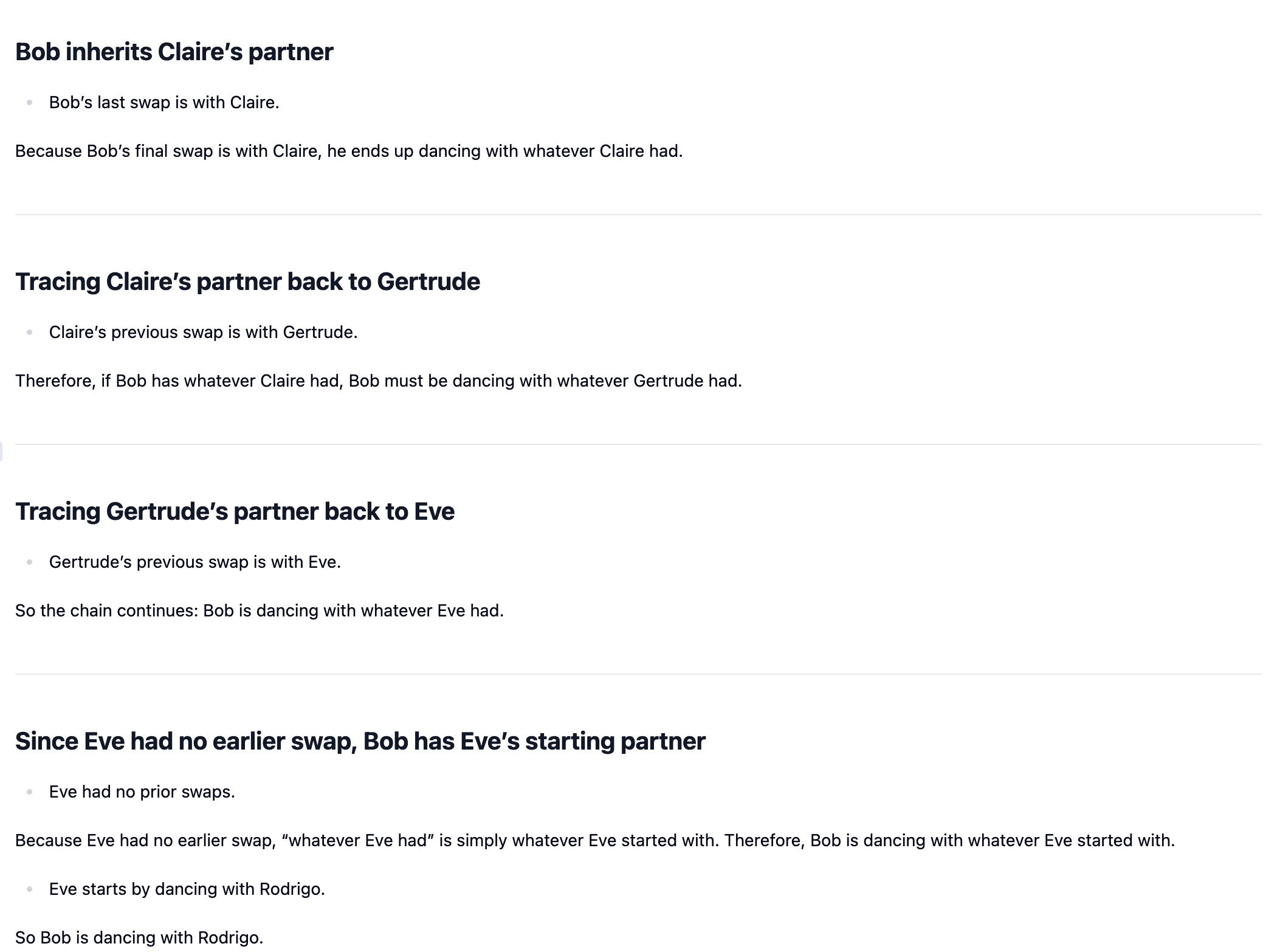}
    \caption{BBH $\times$ Text}
    \label{fig:screenshot_bbh_text}
  \end{subfigure}
  \begin{subfigure}[b]{0.48\textwidth}
    \includegraphics[width=\linewidth]{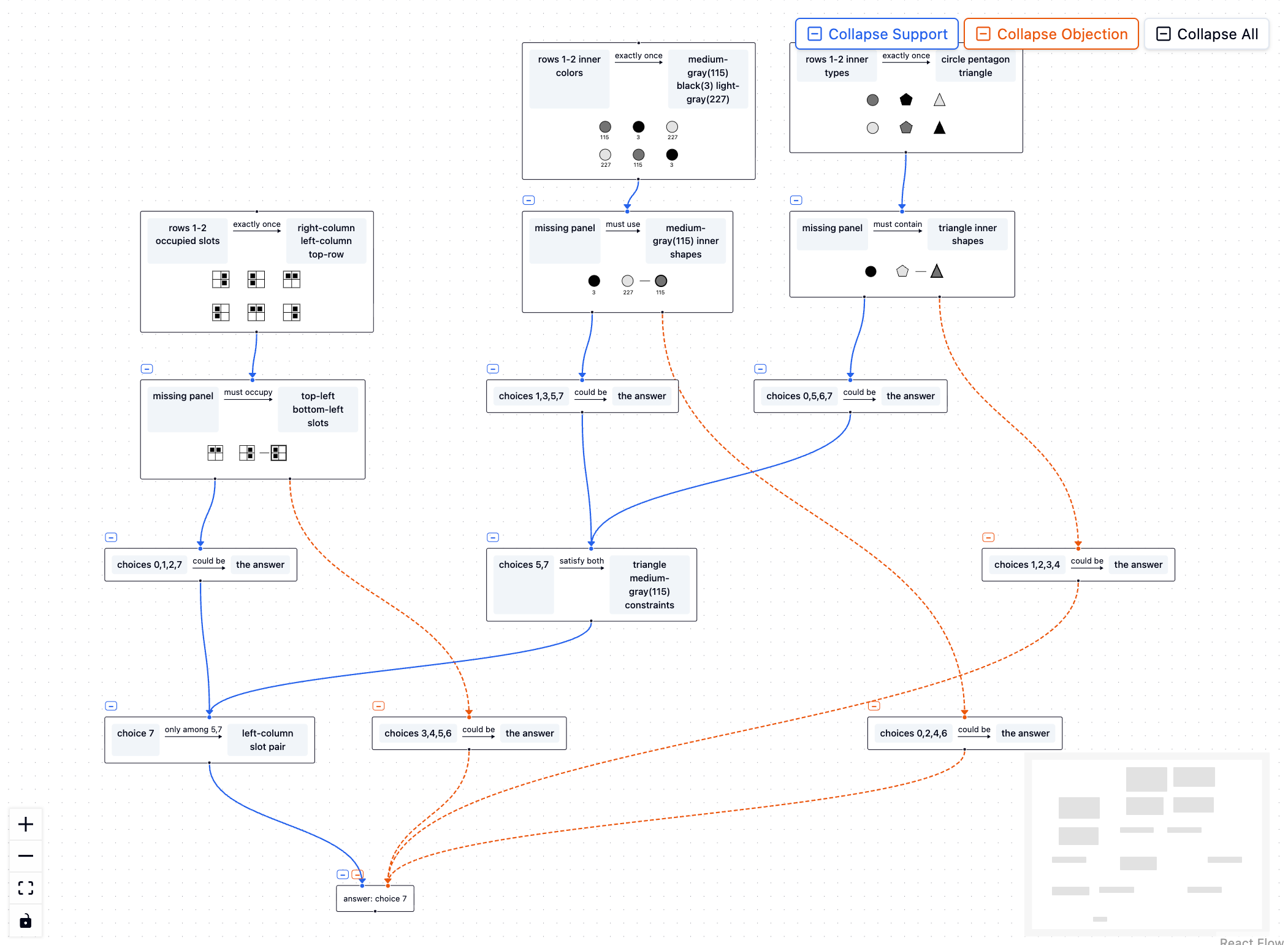}
    \caption{I-RAVEN $\times$ Graph}
    \label{fig:screenshot_iraven_graph}
  \end{subfigure}
  \begin{subfigure}[b]{0.48\textwidth}
    \includegraphics[width=\linewidth]{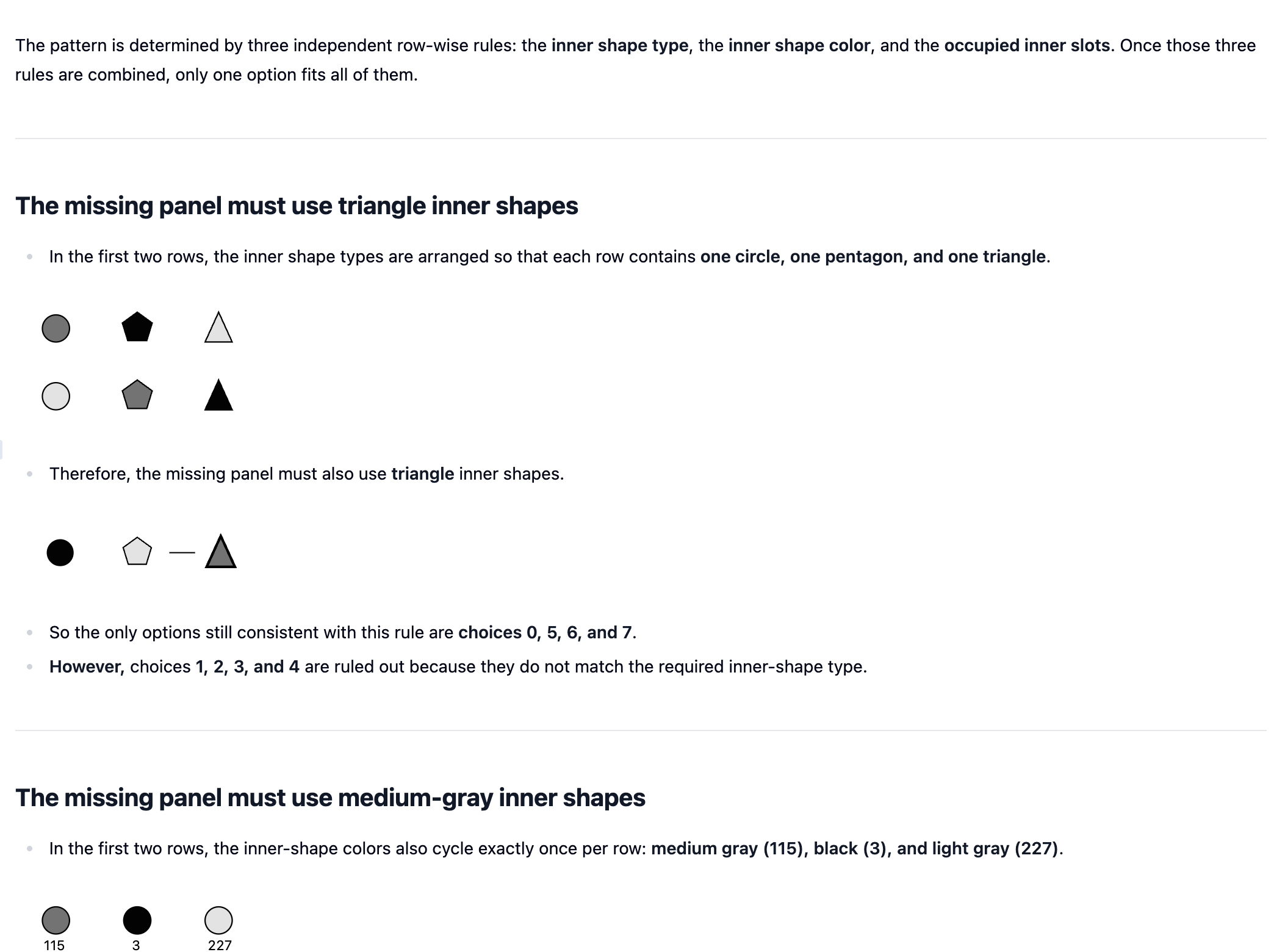}
    \caption{I-RAVEN $\times$ Text}
    \label{fig:screenshot_iraven_text}
  \end{subfigure}

  \caption{Screenshots of the four user study conditions, organized by task modality (verbal vs.\ visual reasoning) and rationale format (graphical vs.\ textual).}
  \Description{Four side-by-side screenshots showing the two rationale formats across the two task modalities, forming a 2×2 grid. (a) BBH × Graph: a directed argument graph for a verbal reasoning task (dancing-partner chain). Oval nodes contain compressed reasoning steps; blue edges indicate supporting relations and orange dashed edges indicate objections, converging to a final answer node. (b) BBH × Text: a textual rationale for the same verbal reasoning task, structured as a sequence of bold section headers (e.g., ``Bob inherits Claire's partner'', ``Tracing Claire's partner back to Gertrude'') each followed by bullet-point premises and a short concluding sentence. (c) I-RAVEN × Graph: a directed argument graph for a visual reasoning task. Nodes contain small shape icons (circles, pentagons, triangles in varying fills) representing pattern observations; blue and orange dashed edges connect nodes across multiple reasoning branches, leading to a final answer node. (d) I-RAVEN × Text: a textual rationale for the same visual reasoning task, with bold headers (e.g., ``The missing panel must use triangle inner shapes'', ``The missing panel must use medium-gray inner shapes'') and bullet points that include embedded shape icons illustrating the pattern rules.}
  \label{fig:user_study_screenshot}
\end{figure*}

We conducted a user study to compare the effects of graphical and textual LLM rationales on human decision making across tasks. Specifically, we investigate:

\begin{itemize}
    \item \textbf{Q1:} How does representing LLM rationales as graphs affect users' interaction experience?
    \item \textbf{Q2:} How does graphical representation affect users' decision-making outcomes?
    \item \textbf{Q3:} Which design features of graphical rationales are helpful, and how can they be improved?
\end{itemize}

\subsection{Conditions}
\label{sec:condition}

We compared graphical LLM rationales against a textual baseline. \rvs{Both conditions were derived from the same underlying argument map.} The textual rationale contained the same information as the full-text content in the graphical rationale's nodes\rvs{, so the comparison isolates presentation format rather than reasoning content}. Headings and bullet points were added so that the textual baseline presented information in a similarly organized form \rvs{and resembled common LLM chat responses, which are now often organized into subsections}. Prompts for textual rationale generation are in Appendix~\ref{ap:user_prompt_text}.

Beyond rationale format (graphical vs.\ textual), we varied task modality and question difficulty to account for different decision-making contexts~\cite{pang_interactive_2025}. We conjectured that the effectiveness of graphical rationales may depend on whether the underlying reasoning is primarily verbal or visual. In verbal reasoning tasks, graphical representations can externalize relational structures that would otherwise require parsing long text. In visual reasoning tasks, which already involve visual patterns, graphical explanations may provide less additional benefit.
We also varied question difficulty, as prior work~\cite{vasconcelos_explanations_2023} suggests users rely more on AI explanations when tasks are sufficiently challenging.
We selected BIG-Bench Hard (BBH)~\cite{suzgun_challenging_2023} as the verbal reasoning task and I-RAVEN~\cite{hu_stratified_2021,yu_cwhyi-raven_2025} as the visual reasoning task. Both datasets report human performance across subcategories, allowing us to select both easy and hard questions. Datasets are described in Section~\ref{sec:dataset}.

Overall, we employed a $2 \times 2$ between-subjects design with (1) rationale format (graphical vs.\ textual) and (2) task modality (verbal (BBH) vs.\ visual reasoning (I-RAVEN)). Within each condition, participants completed both easy and hard questions. Screenshots of the four conditions are shown in \autoref{fig:user_study_screenshot}.

\subsection{Participants}

We conducted a power analysis using G*Power~\cite{faul_statistical_2009} ($f = 0.25$, $\alpha = 0.05$, power $= 0.9$), resulting in a required sample size of 171 participants across four conditions.
After IRB approval, we recruited participants from Prolific using standard quality filters (U.S.-based, $\geqslant$99\% approval rate with $\geqslant$1,000 submissions, native English speakers, desktop users).
Participants received \$4.50 base compensation, with a performance-based bonus of up to \$4.00 if they achieved an overall accuracy above 70\% (\$0.20 per correct answer; \$0.70 for two designated trials in which the AI provided incorrect answers).

Out of 223 participants recruited, 204 valid participants passed the attention checks (BBH $\times$ Graph: 57, BBH $\times$ Text: 50, I-RAVEN $\times$ Graph: 51, I-RAVEN $\times$ Text: 46). 
The final sample was 52\% female ($M_{\text{age}} = 41$), with 27.0\% holding a graduate degree or higher, 45.6\% a bachelor's degree, 17.2\% some college education, 9.3\% a high school diploma, and 1.0\% some high school education or lower. 
Regarding AI experience, 8.8\% had professional or research experience, 54.4\% were regular users, 35.8\% had basic familiarity, and 1.0\% had no prior experience. 
The study lasted 47 minutes on average, yielding an average hourly compensation of $\sim$\$8.85.

\subsection{Study Procedure}

Participants were randomly assigned to one of the four conditions. The study consisted of four stages: pre-study survey, tutorial, task phase, and post-study survey.

In the pre-study survey, participants reported demographics, AI experience~\cite{ma_towards_2025}, AI reliance~\cite{li_reasongraph_2025,scholz_measuring_2025}, and need for cognition~\cite{bucinca_trust_2021}. 
A tutorial then walked them through example questions step by step, introduced the interface, and explained how to interpret the assigned LLM rationales.
During the task phase, participants completed 15 trials. Each trial followed the sequence: viewing question, providing initial answer, viewing LLM rationale, and making revised decision. The system provided rationales only, without explicit answer recommendations, as prior work shows users engage in more analytical reasoning without recommendations~\cite{gajos_people_2022,bucinca_trust_2021}. Rationales were pre-generated to ensure consistency across participants. 

The 15 trials included 5 easy and 10 hard questions. To examine whether rationales help users identify incorrect reasoning, 5 questions (1 easy, 4 hard) were paired with incorrect LLM answers and selected to reflect the error pattern distribution identified from our technical evaluation (Section~\ref{sec:tech_evaluation_finding}).

After completing all trials, participants completed a post-study survey. Attention-check questions were included in both the main study and the post-study survey.

\subsection{Measurements}

\subsubsection{Independent Variables}

Our study included three independent variables:  \textbf{rationale format} (graphical vs.\ textual; between-subjects), \textbf{task modality} (verbal vs.\ visual reasoning; between-subjects), and \textbf{question difficulty} (easy vs.\ hard; within-subjects).

We also measured \textbf{Need for Cognition (NFC)} as an individual-difference variable using a 4-item scale~\cite{cacioppo_efficient_1984,bucinca_trust_2021}, which assesses dispositional tendency to engage in effortful thinking. NFC was included as a moderator in secondary analyses.

\subsubsection{Dependent Variables}

\paragraph{Objective Behavioral Measures}

Our behavioral analysis examines whether participants’ final decisions appropriately track the correctness of AI recommendations. We operationalize trust calibration using four trial-level metrics:

\begin{itemize}
    \item \textbf{Appropriate Trust:} 
    $\frac{\#(\text{accept correct} \,\cup\, \text{reject incorrect})}{\#(\text{all trials})}$

    \item \textbf{Over-Trust:} 
    $\frac{\#(\text{follow incorrect})}{\#(\text{AI incorrect trials})}$

    \item \textbf{Under-Trust:} 
    $\frac{\#(\text{reject correct})}{\#(\text{AI correct trials})}$

    \item \textbf{Error Correction:} 
    $\frac{\#(\text{override incorrect and choose correct})}{\#(\text{AI incorrect trials})}$
\end{itemize}

where $\#(\cdot)$ denotes the number of trials satisfying the condition.

\paragraph{Subjective Self-Report Measures}

After completing all trials, participants rated their experience on nine dimensions: 
Usability~\cite{lewis_umux-lite_2013},
Task Load~\cite{wang_effects_2022,ma_towards_2025},
Satisfaction~\cite{ma_towards_2025},
Engagement\cite{obrien_development_2010},
Helpfulness~\cite{ma_towards_2025,wang_less_2025},
Understanding~\cite{ma_towards_2025,wang_less_2025,wang_effects_2022},
Trust~\cite{ma_towards_2025,wang_less_2025,wang_effects_2022},
Critical Thinking, 
Learning Gain~\cite{vogt_student_2005}.
Participants in the graphical rationale condition additionally rated four design features (Alignment with Human Decision Logic, Semantic Color-Coding, Text Compression, Progressive Disclosure) on a 7-point helpfulness scale~\cite{wang_xagen_2026}.
Two open-ended questions were included in the post-study survey to capture participants’ perceptions and improvement suggestions.
All instruments are in Appendix~\ref{ap:user_presurvey} and \ref{ap:user_postsurvey}.

\subsection{Analysis Method}

For trial-level objective outcomes, we fitted linear mixed-effects models with participant as a random intercept. All categorical predictors were sum-coded ($-1$, $+1$), and models were estimated using maximum likelihood. We computed estimated marginal means (EMMs) and conducted pairwise comparisons.

Subjective outcomes were analyzed at the participant level using $2 \times 2$ between-subjects ANOVA, with within-dataset simple effects examined using Welch's $t$-tests.

We conducted an exploratory moderation analysis on NFC. \rvs{We treated standardized NFC as a continuous moderator by extending the models to include NFC and its interactions.
For consistency with prior work that stratifies participants by NFC~\cite{bucinca_trust_2021}, we additionally report a descriptive median-split analysis.} NFC was median-split within each dataset ($\mathit{Mdn} = 4.00$ for both BBH and I-RAVEN\rvs{, which was also the scale midpoint}) to form Low-NFC and High-NFC subgroups, and models were re-estimated within each subgroup.

Open-ended responses were analyzed using thematic analysis~\cite{braun_using_2006}, separately by task modality and rationale format.

\section{Results}
\subsection{Objective Outcomes}

\begin{figure}[t]
  \centering
  \begin{subfigure}[t]{0.49\linewidth}
    \centering
    \includegraphics[width=\linewidth]{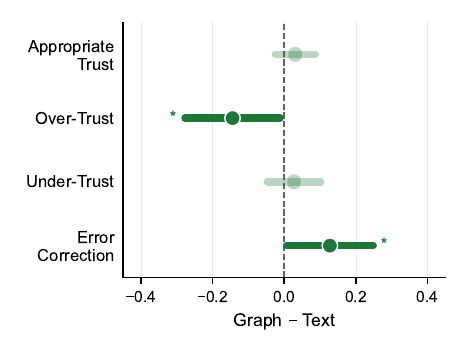}
    \caption{BBH -- Objective (all)}
    \label{fig:bbh_obj}
  \end{subfigure}
  \begin{subfigure}[t]{0.49\linewidth}
    \centering
    \includegraphics[width=\linewidth]{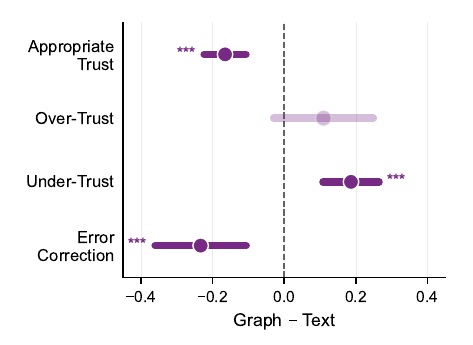}
    \caption{I-RAVEN -- Objective (all)}
    \label{fig:iraven_obj}
  \end{subfigure}
  \begin{subfigure}[t]{0.49\linewidth}
    \centering
    \includegraphics[width=\linewidth]{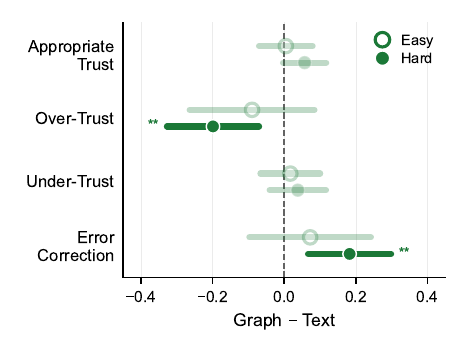}
    \caption{BBH -- Objective
    \\\centering(by difficulty)}
    \label{fig:bbh_obj_diff}
  \end{subfigure}
  \begin{subfigure}[t]{0.49\linewidth}
    \centering
    \includegraphics[width=\linewidth]{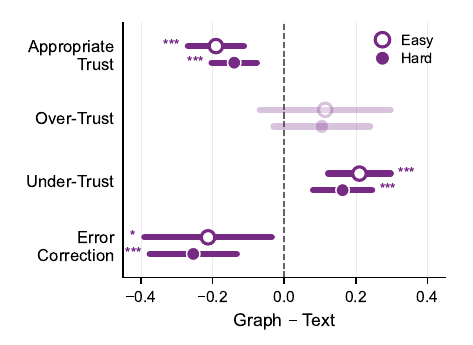}
    \caption{I-RAVEN -- Objective\\
    \centering(by difficulty)}
    \label{fig:iraven_obj_diff}
  \end{subfigure}

  \caption{%
    Objective trust-calibration outcomes (Graph $-$ Text) across task modalities.
    \textbf{Left column:} Estimated Marginal Means collapsed across difficulty.
    \textbf{Right column:} Difficulty-stratified contrasts (open = Easy; filled = Hard).
    In verbal reasoning, the graph advantage is absent on easy questions and emerges on hard ones.
    In visual reasoning, the graph disadvantage is present at both difficulty levels.
    Faded items are non-significant ($p \geq .05$);
    * $p<.05$, ** $p<.01$, *** $p<.001$.
  }
  \Description{Four forest plots arranged in a 2×2 grid, showing estimated marginal means (Graph minus Text contrasts) for four objective trust-calibration outcomes: Appropriate Trust, Over-Trust, Under-Trust, and Error Correction. Each panel plots contrast estimates with 95\% confidence intervals on a horizontal axis ranging from $-0.4$ to $0.4$, with a dashed vertical line at zero. Faded markers indicate non-significant effects ($p \geq .05$); filled markers indicate significant effects with asterisks (* $p<.05$, ** $p<.01$, *** $p<.001$). 
  (a) BBH -- Objective (all): green forest plot, all four outcomes collapsed across difficulty. Over-Trust shows a negative estimate with a single asterisk. Error Correction, which shows a positive estimate with a single asterisk. All other contrasts are near zero and non-significant. 
  (b) I-RAVEN -- Objective (all): purple forest plot, all four outcomes collapsed across difficulty. Appropriate Trust shows a negative contrast (***), Under-Trust a positive contrast (***), and Error Correction a negative contrast (***); Over-Trust is non-significant. 
  (c) BBH -- Objective (by difficulty): green forest plot with open circles for Easy and filled circles for Hard. Over-Trust shows a significant negative contrast for Hard only (**); Error Correction shows a significant positive contrast for Hard only (**); remaining outcomes are non-significant at both difficulty levels. 
  (d) I-RAVEN -- Objective (by difficulty): purple forest plot with open circles for Easy and filled circles for Hard. Appropriate Trust, Under-Trust, and Error Correction are significantly negative at both levels; Over-Trust remains non-significant.}
  \label{fig:obj-combined}
\end{figure}

\paragraph{The rationale format effect reverses between verbal and visual reasoning.}
The rationale format $\times$ task modality interaction was significant across
all four trust-calibration metrics
(Appropriate Trust: $b = 0.049$, $p < .001$;
 Over-Trust: $b = -0.064$, $p = .005$;
 Under-Trust: $b = -0.040$, $p = .004$;
 Error Correction: $b = 0.090$, $p < .001$). The direction of the format effect reverses between
verbal (BBH) and visual (I-RAVEN) reasoning tasks.
Point estimates and 95\% CIs are shown in
\autoref{fig:obj-combined}.

Question difficulty does not explain this reversal: none of the rationale
format $\times$ task modality $\times$ question difficulty three-way interactions
reached significance.
However, difficulty modulates the \emph{magnitude} of format effects:
a marginal format $\times$ difficulty interaction ($b = -0.013$, $p = .068$ for
Appropriate Trust) and a significant task modality $\times$ difficulty interaction
($p < .001$ for all metrics) indicate that the pattern strengthens
on harder items within verbal reasoning.

\paragraph{Verbal reasoning (BBH): graph advantage concentrates on hard questions.}
Collapsed across difficulty, graphical rationales significantly reduced
Over-Trust ($\Delta = -0.144$, $p = .033$)
and increased Error Correction ($\Delta = +0.127$, $p = .040$).
Difficulty-stratified contrasts (\autoref{fig:bbh_obj_diff}) show this benefit is
absent on easy questions and emerges on hard ones:
Over-Trust ($\Delta = -0.199$, $p = .002$)
and Error Correction ($\Delta = +0.182$, $p = .002$)
both reached significance, with Appropriate Trust showing a
marginal positive trend ($\Delta = +0.057$, $p = .064$). This suggests that
explicit argument structure helps only when reasoning demands are
sufficiently high.

\paragraph{Visual reasoning (I-RAVEN): graph disadvantage is robust across difficulty.}
Graphical rationales impaired trust calibration in visual reasoning at both difficulty levels.
Collapsed, Appropriate Trust ($\Delta = -0.165$, $p < .001$),
Under-Trust ($\Delta = +0.186$, $p < .001$),
and Error Correction ($\Delta = -0.233$, $p < .001$) all differed significantly.
This pattern held at both difficulty levels (\autoref{fig:iraven_obj_diff}).
The fact that graphical rationales hurt performance even on simple visual items
rules out a question-difficulty explanation and instead points to a modality mismatch: the graph structure does not provide the perceptual
grounding required for visual reasoning verification, regardless of task demand.

Task modality thus determines the \emph{direction} of the format effect;
question difficulty shapes its \emph{magnitude} within BBH (verbal reasoning) but does not alter the cross-modality reversal.

\subsection{Subjective Experience Diverges from Objective Outcomes}

\begin{figure}[t]
  \centering
  \begin{subfigure}[t]{0.49\linewidth}
    \centering
    \includegraphics[width=\linewidth]{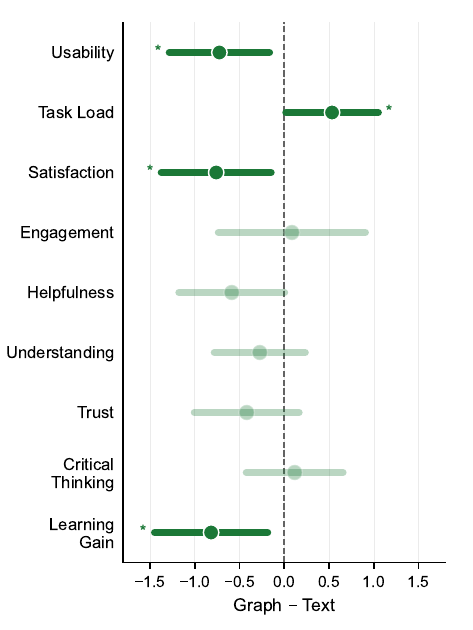}
    \caption{BBH -- Subjective}
    \label{fig:bbh_subj}
  \end{subfigure}
  \begin{subfigure}[t]{0.49\linewidth}
    \centering
    \includegraphics[width=\linewidth]{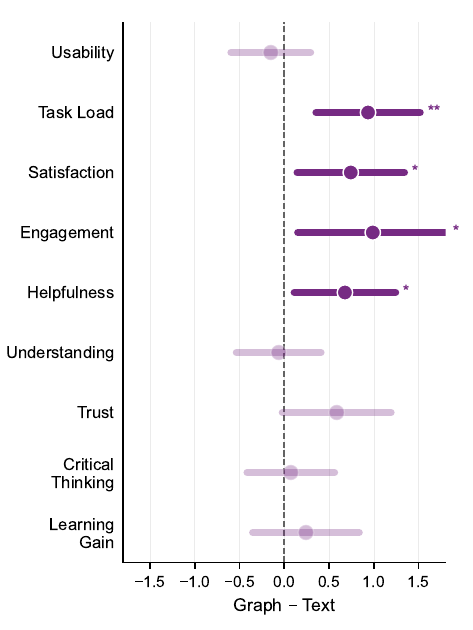}
    \caption{IRAVEN -- Subjective}
    \label{fig:iraven_subj}
  \end{subfigure}
  \caption{%
    Subjective self-report outcomes (Graph $-$ Text) within each task modality.
    Welch's $t$-test simple effects.
    Faded = non-significant ($p \geq .05$);
    * $p<.05$, ** $p<.01$, *** $p<.001$.
  }
  \label{fig:subj-effects}
  \Description{Two forest plots side by side, showing Graph minus Text contrasts for nine subjective self-report outcomes within each task modality. Each panel plots contrast estimates with 95\% confidence intervals on a horizontal axis ranging from $-1.5$ to $1.5$, with a dashed vertical line at zero. Faded markers indicate non-significant effects ($p \geq .05$); filled markers indicate significant effects with asterisks (* $p<.05$, ** $p<.01$, *** $p<.001$). Outcomes listed from top to bottom are: Usability, Task Load, Satisfaction, Engagement, Helpfulness, Understanding, Trust, Critical Thinking, and Learning Gain. (a) BBH -- Subjective: green forest plot. Usability shows a negative contrast (*); Task Load shows a positive contrast (*); Satisfaction shows a negative contrast (*); Learning Gain shows a negative contrast (*). The remaining five outcomes -- Engagement, Helpfulness, Understanding, Trust, and Critical Thinking -- are non-significant and faded. (b) I-RAVEN -- Subjective: purple forest plot. Task Load shows a positive contrast (**); Satisfaction shows a positive contrast (*); Engagement shows a positive contrast (*); Helpfulness shows a positive contrast (*). The remaining five outcomes -- Usability, Understanding, Trust, Critical Thinking, and Learning Gain -- are non-significant and faded.}
\end{figure}

Despite the split in objective outcomes, subjective ratings show the
opposite pattern (\autoref{fig:subj-effects}),
revealing a dissociation between how participants
\emph{experienced} graphical rationales and how they \emph{performed} with them.

\paragraph{Verbal reasoning (BBH): graph works better but feels worse.}
Where graphical rationales objectively reduced over-trust and
improved error correction, participants rated the experience
lower on Satisfaction ($\Delta = -0.76$, $p = .017$),
Usability ($\Delta = -0.72$, $p = .013$),
and Learning Gain ($\Delta = -0.82$, $p = .013$),
and reported higher Task Load ($\Delta = +0.53$, $p = .045$).

\paragraph{Visual reasoning (I-RAVEN): graph feels better but works worse.}
Where graphical rationales objectively impaired trust calibration,
participants reported greater Satisfaction
($\Delta = +0.74$, $p = .017$), Engagement
($\Delta = +0.99$, $p = .023$), and Helpfulness
($\Delta = +0.68$, $p = .021$), alongside elevated
Task Load ($\Delta = +0.94$, $p = .002$).

This pattern reveals that graphical
rationales feel engaging in visual tasks---where they most likely produce
miscalibrated trust, while imposing friction in verbal tasks---where they
deliver the greatest objective benefit.

\subsection{\rvs{Need for Cognition Moderates the Trust-Calibration Effect}}
\label{sec:results-nfc}

We examined whether Need for Cognition moderated the format effect. \rvs{The rationale format $\times$ NFC interaction was significant for Over-Trust ($b = -0.072$, $p = .009$) and Under-Trust ($b = +0.035$, $p = .032$), but not for Appropriate Trust or Error Correction. Higher-NFC participants showed a stronger reduction in Over-Trust but also a stronger increase in Under-Trust with graphical rationales. No significant rationale format $\times$ NFC interactions were found for any subjective outcome. }

\begin{figure}[t]
  \centering
  \begin{subfigure}[t]{0.49\linewidth}
    \centering
    \includegraphics[width=\linewidth]{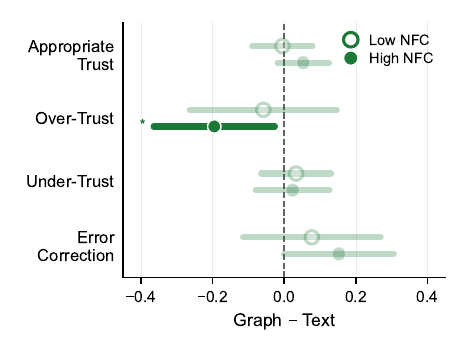}
    \caption{BBH -- Objective}
    \label{fig:bbh_obj_nfc}
  \end{subfigure}
  \begin{subfigure}[t]{0.49\linewidth}
    \centering
    \includegraphics[width=\linewidth]{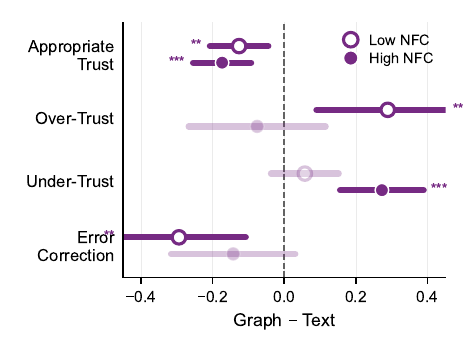}
    \caption{I-RAVEN -- Objective}
    \label{fig:iraven_obj_nfc}
  \end{subfigure}

  \vspace{0.4em}

  \begin{subfigure}[t]{0.495\linewidth}
    \centering
    \includegraphics[width=\linewidth]{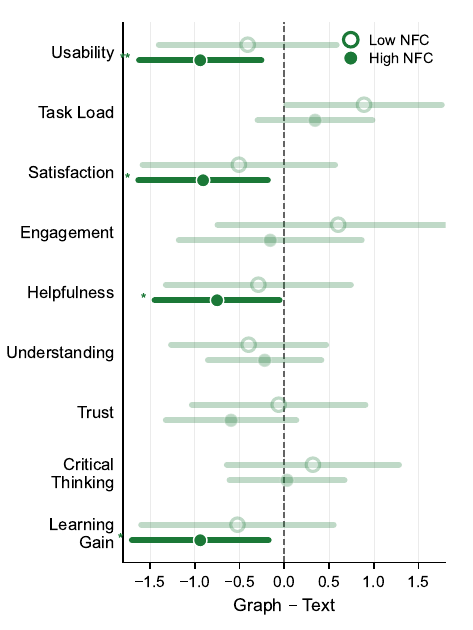}
    \caption{BBH -- Subjective}
    \label{fig:bbh_subj_nfc}
  \end{subfigure}
  \begin{subfigure}[t]{0.495\linewidth}
    \centering
    \includegraphics[width=\linewidth]{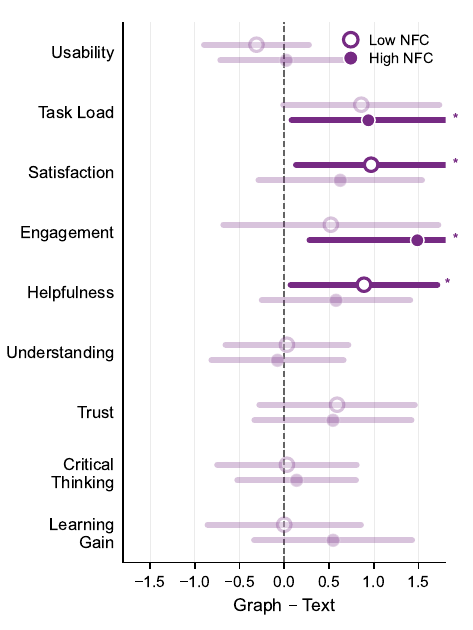}
    \caption{I-RAVEN-- Subjective}
    \label{fig:iraven_subj_nfc}
  \end{subfigure}

  \caption{%
    NFC-stratified format effects (Graph $-$ Text), median-split ($\mathit{Mdn} = 4.00$).
    Open = Low NFC; filled = High NFC.
    \textbf{Top:} Objective outcomes. \textbf{Bottom:} Subjective outcomes.
    In verbal reasoning, the objective benefit concentrates in High-NFC participants;
    in visual reasoning, calibration impairment is present across both NFC groups.
    Faded = non-significant; * $p<.05$, ** $p<.01$, *** $p<.001$.
  }
  \Description{Four forest plots arranged in a 2×2 grid, showing Graph minus Text contrasts stratified by Need for Cognition (NFC), median-split at $Mdn = 4.00$. Open circles represent Low NFC participants; filled circles represent High NFC participants. Each panel plots contrast estimates with 95\% confidence intervals on a horizontal axis, with a dashed vertical line at zero. Faded markers indicate non-significant effects ($p \geq .05$); asterisks denote significance (* $p<.05$, ** $p<.01$, *** $p<.001$). (a) BBH -- Objective: green forest plot with four outcomes (Appropriate Trust, Over-Trust, Under-Trust, Error Correction). Over-Trust shows a significant negative contrast for High NFC only; Error Correction shows a marginal positive contrast for High NFC only; all other contrasts are non-significant for both NFC groups. (b) I-RAVEN -- Objective: purple forest plot with the same four outcomes. Appropriate Trust is significantly negative for both Low and High NFC (***); Under-Trust is significantly positive for High NFC (***); Over-Trust shows a significant positive contrast for Low NFC (**); Error Correction shows a significant negative contrast for Low NFC (**). (c) BBH -- Subjective: green forest plot with nine outcomes (Usability, Task Load, Satisfaction, Engagement, Helpfulness, Understanding, Trust, Critical Thinking, Learning Gain) on an axis from $-1.5$ to $1.5$. Usability, Satisfaction, Helpfulness and Learning Gain show significant negative contrasts for High NFC; other outcomes are non-significant for both NFC groups. (d) I-RAVEN -- Subjective: purple forest plot with the same nine outcomes. Helpfulness shows a significant positive contrast for Low NFC (*); Engagement shows a significant positive contrast for High NFC; Satisfaction shows a significant positive contrast for Low NFC; Task Load shows a significant positive contrast for High NFC; remaining outcomes are non-significant for both groups.}
  \label{fig:nfc-effects}
\end{figure}

\rvs{To descriptively interpret this moderation, we additionally stratified participants by a median split.} NFC-stratified contrasts are shown in
\autoref{fig:nfc-effects}. In verbal reasoning (BBH), low-NFC participants showed no reliable difference between
rationale formats. High-NFC participants, by contrast, showed reduced
Over-Trust ($\Delta = -0.195$, $p = .024$)
and a marginal improvement in Error Correction
($\Delta = +0.152$, $p = .052$), together with lower subjective ratings on
Usability ($\Delta = -0.938$, $p = .009$),
Satisfaction ($\Delta = -0.906$, $p = .017$),
Learning Gain ($\Delta = -0.938$, $p = .019$),
and Helpfulness ($\Delta = -0.750$, $p = .039$).
In visual reasoning, the graph disadvantage on trust calibration was present across both
NFC subgroups. Both groups also rated graphical rationales more positively.

\rvs{Taken together, the continuous analysis supports NFC moderation of objective trust calibration, and the median-split suggests that the calibration benefit in verbal reasoning concentrates among High-NFC participants. In contrast, subjective effects showed no reliable NFC moderation in the continuous analysis, so the subgroup differences in ratings should be interpreted as exploratory. 
}


\subsection{Design Features and Qualitative Themes}
\label{sec:results-features}

\paragraph{All graphical features were rated as helpful.}
Participants in graphical rationale conditions rated all four design features above the scale midpoint ($p < .001$): Progressive Disclosure ($M = 5.66$),
Semantic Color-Coding ($M = 5.58$), Text Compression ($M = 5.45$),
and Alignment with Human Decision Logic ($M = 5.39$).

\paragraph{Qualitative themes mirror the task modality split.}
Thematic analysis of open-ended responses showed that participants in the two
task modalities used the same design in different ways, despite sharing a common appreciation of its stepwise structure
(BBH: 22/57 = 38.6\%; I-RAVEN: 21/51 = 41.2\%).

In verbal reasoning (BBH), the graph served as a logic-tracing tool whose
structural transparency made AI errors visible and prompted correction,
but at the cost of perceived trustworthiness when errors occurred.
Complaints centered on branching complexity (17.5\% of BBH $\times$ Graph vs.\ 7.8\% of
I-RAVEN $\times$ Graph participants), and a disproportionate share of BBH $\times$ Graph
participants (22.8\% vs.\ 5.9\%) tied their confidence in the design directly
to the AI's reasoning accuracy:
\textit{``Once it gets one thing wrong it's hard to follow the logic.
  Once I encountered that I didn't trust the AI to give me a correct answer''}~(P75).
Improvement requests focused on simplification including fewer nodes and shorter
reasoning chains.

In visual reasoning (I-RAVEN), the graph was valued for decomposing complex visual patterns into trackable attributes:
\textit{``It was very helpful as it breaks down complex visual patterns into
  simple ones''}~(P48).
However, the gap between the graph's abstract structure and the perceptual task limited its usefulness. Participants asked for color-coded attribute tracks, visual legends, and
direct highlighting in the source images.
Even participants in I-RAVEN $\times$ Text condition frequently requested visual
overlays (12/44 = 27.3\%), indicating that the need for perceptual grounding
is broadly felt and not met by either format.


\subsection{Summary}

In summary, graphical rationales improve trust calibration in verbal reasoning (BBH)---particularly for analytically engaged participants and on harder questions---albeit at the cost of perceived usability. 
In contrast, the same design feels more engaging and helpful in visual reasoning (I-RAVEN), yet consistently impairs calibrated trust. 
Qualitative evidence points to a shared underlying mechanism: in verbal reasoning, the graph acts as a traceable logical structure that exposes AI errors, whereas in visual reasoning, it functions as a decomposition aid that is intuitive but lacks sufficient perceptual support for reliable verification.

\section{Discussion}
Our results show that graphical rationales \rvs{did not help uniformly: they} improved trust calibration for verbal reasoning
yet impaired it for visual reasoning, while subjective ratings showed the
opposite pattern in each domain.
We discuss the implications of these findings in this section.

\subsection{Modality Mismatch}
\label{sec:discuss-modality}

The reversal in effectiveness across task modalities can be explained by the complementarity of representations~\cite{sweller_cognitive_1998,larkin_why_1987}. A diagram helps when its structure matches the structure of the problem. It allows problem-related operations to be done perceptually rather than symbolically, and reduces the working memory load~\cite{larkin_why_1987}.

In verbal reasoning, the argument graph achieves this match---each node maps to a logical inference and edges encode dependencies that participants can inspect sequentially.
But visual reasoning operates under different representational demands. I-RAVEN tasks require integrating spatial and figural information across multiple pattern regularities. The graphical rationale, however, represents this process through an abstract argument graph disconnected from the source images. Rather than grounding participants in the perceptual features, it asks them to re-encode visual information into propositional nodes, imposing extraneous cognitive load~\cite{mayer_nine_2003,sweller_cognitive_1998}. Participants must maintain the spatial matrix in working memory while also
parsing the abstract graph---a dual-channel demand that, under multiple-resource
theory~\cite{wickens_multiple_2002}, degrades rather than supports verification.
Qualitative evidence supports this: I-RAVEN participants specifically requested visual overlays and color-coded attribute tracks.
Importantly, this mismatch is not explained by task difficulty. The reversal held at both difficulty levels and the three-way interaction was non-significant. This points to a structural incompatibility between the graph's abstract vocabulary and the verification demands of visual reasoning.

\rvs{However, I-RAVEN tests a specific instantiation of graphical rationales, i.e., argument-map-style graphs, rather than graphical rationales for visual reasoning in general. The poor calibration outcomes should not be interpreted as evidence that graphical rationales are universally harmful for visual tasks. Instead, they reveal an important \textit{boundary condition of our design}: abstract argument graphs are poorly matched to tasks requiring perceptual information from source images. Argument-map-style graphs appear better suited to verbal reasoning, where nodes and edges can represent logical dependencies. In contrast, visual reasoning likely requires explanations perceptually grounded in source stimuli.
}

\subsection{The Objective–Subjective Dissociation}
\label{sec:discuss-dissociation}

In verbal reasoning, graphical rationales improved calibration but were rated as more demanding and less satisfying. In visual reasoning, they impaired calibration but were rated as more engaging and helpful. This shows that participants' judgments of their own reasoning quality are unreliable when rationale format varies.

\rvs{Response time offers complementary behavioral evidence. 
We compared per-trial response time and found no format effect in verbal reasoning, whereas in visual reasoning participants spent less time with graphical than with textual rationales (28.2s vs. 36.7s, $p = .013$). This pattern further reinforces the objective--subjective dissociation.
In BBH, graphical rationales increased perceived task load without increasing time spent; in I-RAVEN, they were rated as more demanding yet processed more quickly.
}

\rvs{Therefore, }for visual reasoning, the structured graph created a sense of comprehension that substituted for genuine verification---an effect similar to the illusion of explanatory depth~\cite{rozenblit_misunderstood_2002}. \rvs{The risk that graphical presentation gives flawed reasoning unwarranted credibility was realized here.} For verbal reasoning, the graph made AI errors highly visible, improving calibration but reducing perceived trustworthiness. This is consistent with prior work showing that increased model transparency can reduce user confidence even when it improves decision quality~\cite{poursabzi-sangdeh_manipulating_2021,bucinca_trust_2021}.
This tension reflects a core challenge for explainable AI design: formats that support calibrated trust may undermine the subjective experience that drives continued engagement.
\rvs{Graph structure thus can support calibration when its representation matches the task, but may increase the perceived credibility of flawed rationales when it does not.}

\subsection{Analytic Engagement as a Gating Condition}
\label{sec:discuss-nfc}

In verbal reasoning, only high-NFC participants showed reduced Over-Trust and improved Error Correction with graphical rationales. This suggests graphical rationales function as argument scaffolds whose benefit requires effortful engagement.
This is consistent with the Elaboration Likelihood Model~\cite{cacioppo_efficient_1984}, which posits that central-route processing---careful evaluation of argument quality---requires both ability and motivation. While the argument graph provides structure that facilitates elaboration, effective evaluation still depends on users’ willingness to engage with it.

The graph advantage also emerged exclusively on hard questions, where explicit structure provides scaffolding for deeper inference. 
Prior work on AI-assisted decision-making has similarly shown that AI explanation benefits are largest when task difficulty is high enough~\cite{vasconcelos_explanations_2023}.

Together, these findings indicate that graphical rationales impose an onboarding barrier. They require a threshold of effortful engagement to deliver their benefit.

\subsection{Design Implications}
\label{sec:discuss-design}

\paragraph{Graph Complexity as an Inverted-U Function.}
Verbal-reasoning participants complained about branching complexity, suggesting a non-linear relationship between node count and calibration benefit. Rationale graphs should be pruned to retain only decision-critical nodes. As the highest-rated design feature, progressive disclosure is a promising mechanism for managing this tradeoff by collapsing branches on demand.

\paragraph{Canvas-Based Presentation for Visual Tasks.}
The visual-reasoning findings motivate a different design: rather than imposing abstract argument graphs, rationale interfaces should anchor explanations directly in the source visual representation, preserve perceptual grounding, and incorporate argument structure.

\paragraph{Hybrid Representations.}
Aligning with the redundancy principle in multimedia learning~\cite{mayer_nine_2003}, the modality-split findings motivate research on hybrid rationales that adapt their representational format based on task type, e.g., graph-structured arguments for verbal reasoning and spatial annotations for visual reasoning. We preliminarily explored augmenting verbal reasoning with additional visual cues (e.g., SVG-based spatial diagrams for BBH logical deduction tasks; see Appendix~\ref{ap:system_bbh_svg}). A subset of participants from our formative interviews indicated that embedding such visual cues within graphical rationales largely improved perceived clarity, suggesting a promising direction for future work.

\subsection{Limitations and Future Work}
\label{sec:discuss-limitations}

Several limitations constrain the generalizability of our findings.
First, our study focuses on well-structured QA tasks, whereas real-world reasoning often involves ambiguity and value-laden trade-offs that do not decompose into discrete steps.
Second, we adopt a decision-making paradigm and do not capture more open-ended, subjective aspects of human–AI collaboration, such as whether AI introduces novel perspectives or supports co-creative thinking.
\rvs{Third, our findings are tied to one instantiation of graphical rationales, i.e., argument-map-style graphs with condensed node text. The visual-reasoning results in particular may not generalize to other visual tasks or to other forms of graphical rationale representation.}
 
A key direction for future work is moving from read-only to \emph{editable} rationale graphs, where users can modify reasoning steps and prompt the model to re-evaluate from intermediate states.

\section{Conclusion}
\rvs{We explored when and how graphical LLM rationales support human decision making, using} \systemName, \rvs{a prototype} that transforms linear LLM rationales into interactive argument graph visualizations. Grounded in a formative co-design study, \systemName combines argument mapping with text condensation and adapts graph complexity through progressive disclosure.
A controlled user study ($N = 204$) showed that graphical rationales \rvs{did not help uniformly: they} improved trust calibration for verbal reasoning---reducing over-trust and increasing error correction, especially on hard questions and among analytically engaged users---yet were perceived as more cognitively demanding and less satisfying. For visual reasoning, graphical rationales impaired trust calibration but were rated as more engaging, helpful, and satisfying, which points to a mismatch between the graph's abstract structure and the perceptual demands of the task. \rvs{In each modality, the format that better supported calibrated decisions was thus not the one users preferred.}
Our findings provide \rvs{empirical design knowledge} for next-generation reasoning-aware AI interfaces that adapt rationale presentation to task modality and user needs.

\begin{acks}
We would like to thank our colleagues Vasiliki Charisi, Shuo Sun, and Alok Prakash at Singapore–MIT Alliance for Research and Technology (SMART) center for their thoughtful feedback and support throughout this project.
We are also grateful to all participants who contributed their time to our formative study and user study. Finally, we thank all anonymous reviewers for their insightful and constructive comments. 
This research has been supported by the National Research Foundation (NRF), Prime Minister’s Office, Singapore, under its Campus for Research Excellence and Technological Enterprise (CREATE) program, through the Mens, Manus and Machina (M3S) interdisciplinary research group (IRG) of the Singapore–MIT Alliance for Research and Technology (SMART) center.
\end{acks}

\bibliographystyle{ACM-Reference-Format}
\bibliography{main}

\clearpage
\onecolumn
\appendix
\renewcommand{\thefigure}{\thesection\arabic{figure}}
\setcounter{figure}{0}

\section{Formative Study Details}

\subsection{Prompt for Converting Text to Argument Map}
\label{ap:formative_prompt}

\paragraph{Instruction Prompt}
\begin{Verbatim}[breaklines=true, fontsize=\small]
You are an expert logic analyzer and you are good at making argument map.
\end{Verbatim}

\paragraph{System Prompt}
\begin{Verbatim}[breaklines=true, fontsize=\small]
You are to transform written text into an **argument map** following these steps:
1. **Identify all claims** made by the author.
2. **Rewrite each claim** as an independent, concise statement.
- Keep it **minimal but readable**, just enough for humans to understand.
- Avoid redundancy and filler words.
3. **Classify** each statement as a *premise*, *sub-conclusion*, or *main conclusion*.
4. **Optionally** include implied premises or conclusions if they are necessary for logical completeness.
5. **Represent** each statement as a node (box) in JSON format.
6. **Connect** nodes with directed edges (arrows) that indicate *support* relationships — from premise(s) to sub-conclusion(s) or conclusion(s).

**Edge rules**
- Use arrows only to show *support* (never opposition).
- If multiple premises support the same conclusion, each should have its own edge to that conclusion.
- Sub-conclusions can serve as intermediate nodes supported by earlier premises and supporting later conclusions.
- **Do not** connect sibling premises to each other unless one explicitly depends on the other.
- **Do not** create edges between independent reasons; they should each point directly to the conclusion they support.
\end{Verbatim}

\paragraph{User Input Template}
\begin{Verbatim}[breaklines=true, fontsize=\small]
text: <rationale text>
\end{Verbatim}

\paragraph{Example Input}
\begin{Verbatim}[breaklines=true, fontsize=\small]
text: The model selected "airport" because the phrase "checked baggage" strongly suggests air travel. Checking luggage is a common airport activity.
Therefore, the most plausible answer is "airport." Other options such as "garbage can," "military," "jewelry store," and "safe" are less consistent with the context.
\end{Verbatim}

\paragraph{Structured Output Schema (Using Zod)}
\begin{Verbatim}[fontsize=\small]
{
"nodes": [
  {
    "id": "...",
    "data": {
      "label": "..."
    },
    "type": "premise" | "sub-conclusion" | "conclusion"
  }
],
"edges": [
  {
    "id": "...",
    "source": "...",
    "target": "..."
  }
]
}
\end{Verbatim}


\clearpage
\subsection{Low-Fidelity Interface Screenshot}
\label{ap:formative_screenshot}
\begin{figure}[H]
\centering
  \includegraphics[width=0.91\textwidth]{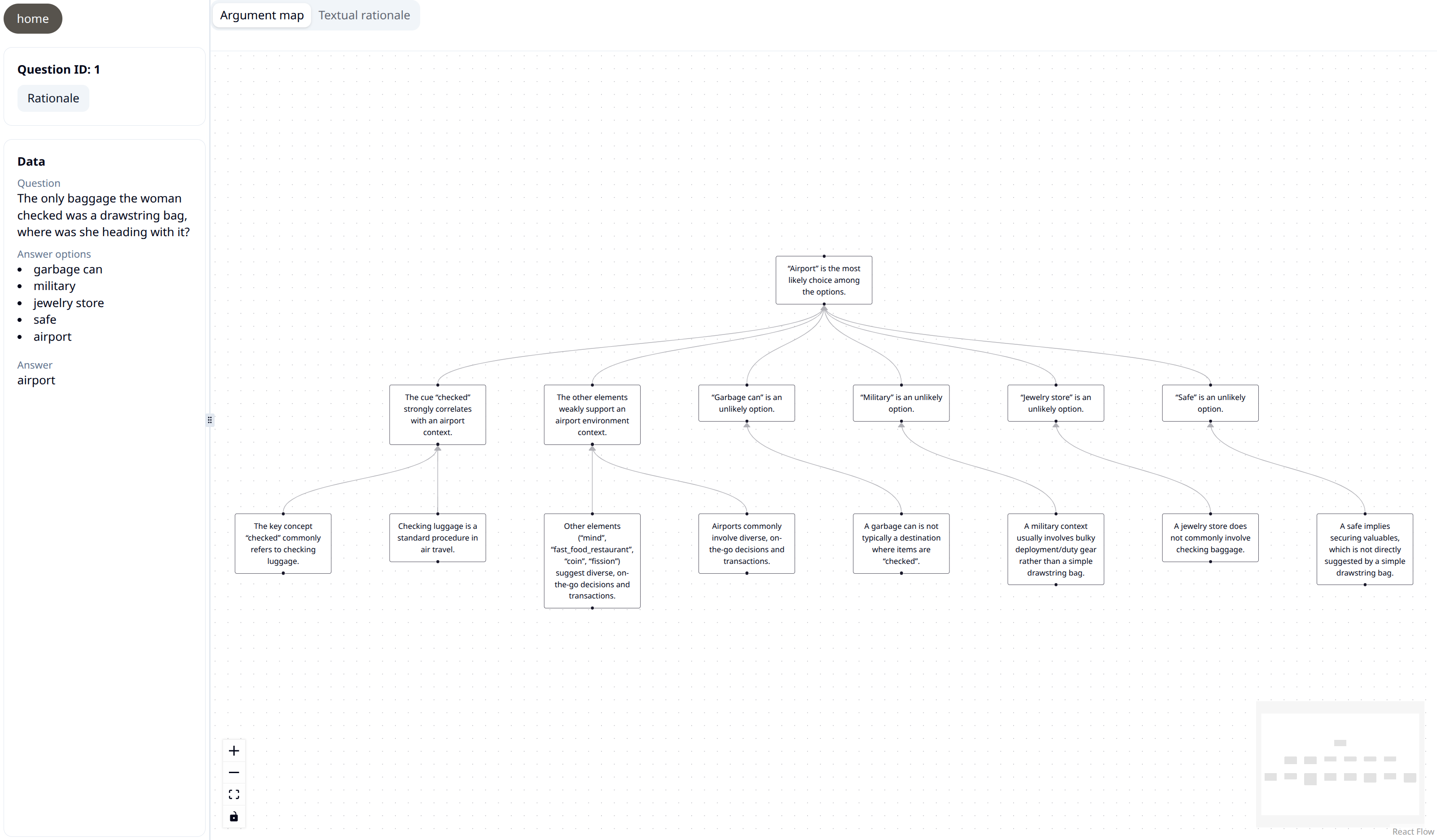}
  \caption{Screenshot of the low-fidelity prototype used in the formative study. The left panel shows a CommonsenseQA question. Participants first reviewed the textual rationale on the right panel before switching to the graphical rationale for comparison.}
\Description{A two-panel web interface. The left panel lists a CommonsenseQA question and multiple-choice answer options, with a ``Rationale'' button. The right panel shows a paragraph of text describing the LLM's reasoning process for selecting ``airport'' as the answer.}
\end{figure}

\subsection{Example Participant Sketches}
\label{ap:formative_sketch}

\begin{figure}[H]
  \centering
  \begin{subfigure}[t]{0.55\linewidth}
    \centering
    \includegraphics[width=\linewidth]{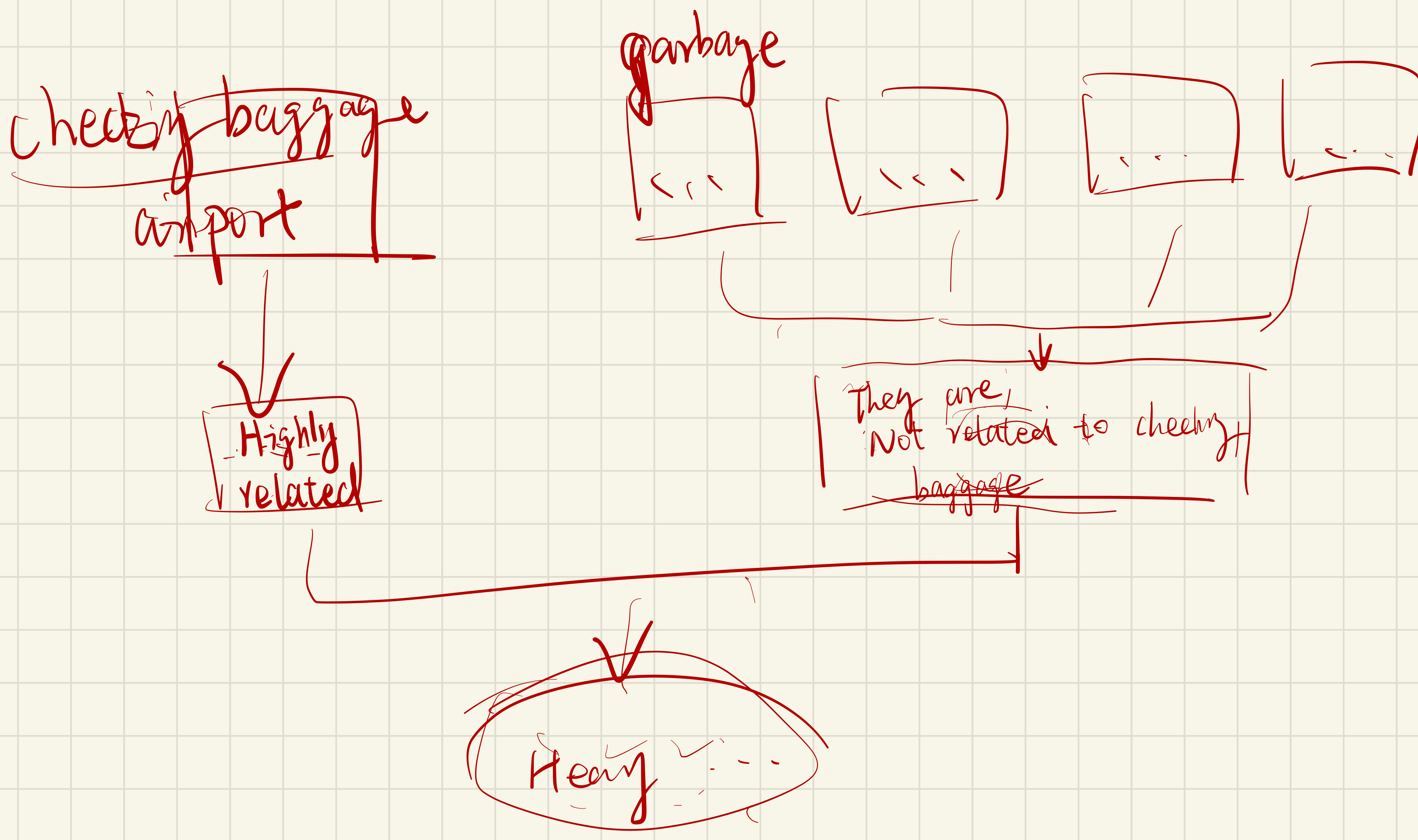}
    \caption{}
  \end{subfigure}
  \hfill
  \begin{subfigure}[t]{0.3\linewidth}
    \centering
    \includegraphics[width=\linewidth]{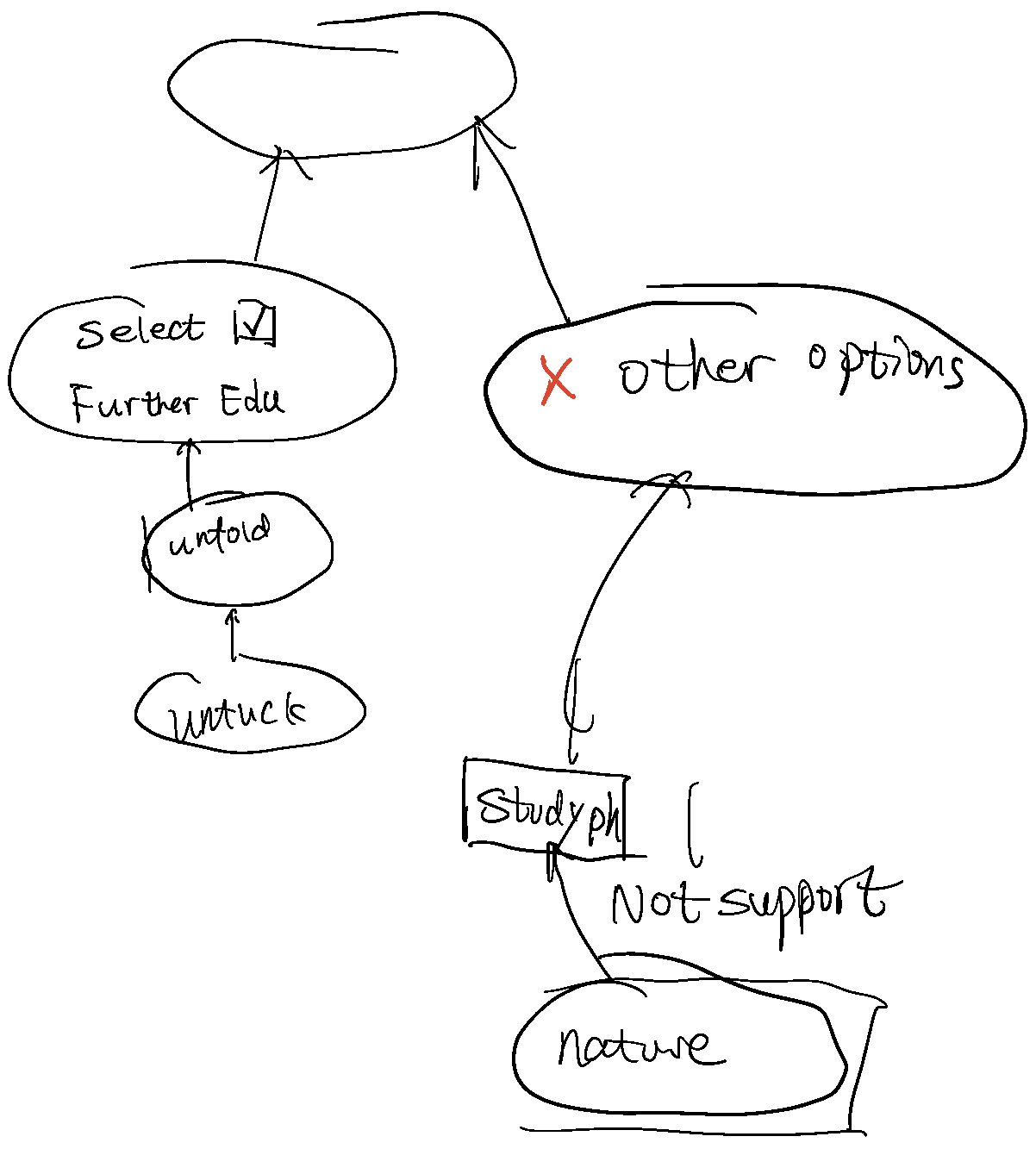}
    \caption{}
  \end{subfigure}
  \caption{Example participant sketches from the formative co-design study. (a) P1 sketched a graph that groups the correct answer (``checking baggage / airport'') and incorrect options (e.g., ``garbage'') into separate branches, with a shared node summarizing why the distractors are unrelated, then leading to a final conclusion node. (b) P10 sketched a graph distinguishing the selected answer (``Further Edu,'' marked with a checkmark) from rejected options (marked with a cross), using hierarchical nodes with progressive unfolding and explicit ``not support'' edges for elimination.
  }
\Description{Two hand-drawn sketches on paper. Left (P1): a rough argument map in red pen showing ``checking baggage / airport'' connecting to a ``highly related'' node on the left branch, and multiple distractor options connecting to a ``they are not related to checking baggage'' node on the right branch, both leading to a conclusion node at the bottom. Right (P10): a cleaner sketch in black pen showing a top conclusion node, with a left branch for the correct answer (``Select Further Edu'' with a checkmark, supported by ``untold'' and ``untuck'' sub-nodes) and a right branch for rejected options (marked with a red X, connected via ``StudyPh'' and a ``Not support'' edge to a ``nature'' node at the bottom).
}

\end{figure}

\clearpage
\section{System Implementation Details}

\subsection{Prompts for Graphical Rationale Generation}
\label{ap:system_prompt}
\subsubsection{Question--Answer Pair to Reasoning Structure}
\label{ap:system_prompt_QA_reasoning}

\paragraph{Developer Prompt}
\begin{Verbatim}[fontsize=\small]
transform a question-answer pair into an **argument map** — a structured reasoning graph.

**node rules**
- each node must express exactly **one atomic claim**. if a sentence covers multiple independent aspects (e.g.,
different attributes, different entities) or chains multiple reasoning steps (e.g., "A, so B, therefore C"),
split it into separate nodes connected by an edge.
- similarly, if one reasoning step resolves multiple independent entities (e.g., "mango → position 5 and kiwi →
position 6"), split them into separate nodes — one per entity.
- each node label must be a **complete sentence** clearly stating the claim or observation.
- classify each node as *premise*, *sub-conclusion*, or *conclusion*.
- the final conclusion node states the chosen answer.
- use the fewest nodes necessary to convey the full reasoning. every node must earn its place — if removing a
node would not lose any logical step, remove it.

**edge rules**
- each edge must have a *relation*: **support** (reason in favor) or **objection** (reason against).
- **support**: use when a premise provides evidence FOR a claim. e.g., premise "inner shape is triangle"
--support--> "choice 1 could be the answer".
- **objection**: use when a premise provides evidence AGAINST a candidate. e.g., premise "these choices do not
show four triangles" --objection--> "choices 2-7 could be the answer".
- **elimination flow**: a candidate sub-conclusion whose set does NOT include the final answer is an eliminated
candidate — it contradicts the conclusion and connects via **objection**. full chain: evidence --objection-->
eliminated candidate --objection--> conclusion.
- **surviving candidate flow**: a candidate sub-conclusion whose set INCLUDES the final answer is narrowing
toward the answer — it connects to the conclusion (or to a more specific surviving candidate) via **support**.
e.g., "choices 2 and 6 fit the size constraint" --support--> "answer: choice 2".
- if multiple premises support or object to the same node, each should have its own edge.
- sub-conclusions may support other sub-conclusions to form deeper reasoning chains when the logic requires
intermediate steps.
- do not connect sibling premises unless one explicitly depends on the other.
\end{Verbatim}

  \paragraph{Output Schema}
  \begin{Verbatim}[fontsize=\small]
  {
    "nodes": [
      {
        "id": "...",
        "data": {
          "label": "...",
          "svg": "..." | null
        },
        "type": "premise" | "sub-conclusion" | "conclusion"
      }
    ],
    "edges": [
      {
        "id": "...",
        "source": "...",
        "target": "...",
        "relation": "support" | "objection"
      }
    ]
  }
  \end{Verbatim}

  \subsubsection{Reasoning Structure to Graphical Rationale}

  \paragraph{Developer Prompt}
  \begin{Verbatim}[fontsize=\small]
  you are given a list of reasoning nodes with full-sentence labels.
  for each node, produce a compressed representation:

  1. try to decompose the label into a **triplet**:
     - subject: the main entity or concept (2-4 keyword tokens)
     - predicate: the relationship or action (1-3 words)
     - object: the related entity, value, or result (2-4 keyword tokens)
     - also provide a short fallback label (max 6 words)
  2. if the sentence does not fit a natural subject-predicate-object structure, set subject/predicate/object to
  null and provide only the label.

  **important**:
  - labels must contain concrete specifics (attribute names, values, choice numbers) — never vague summaries like
  "missing panel specification" or "rule analysis". a good label lets the reader understand the claim without
  hovering for details.
  - when the original label describes a rule or pattern (e.g. progression, constant, distribute three), the
  compressed label must preserve the rule name and key values. e.g. "color progression: 31→59→87", not just "color
  fixed as grayscale 87".
  - preserve human-readable formatting from the original: e.g. keep "grayscale 199" not "199", keep "the largest"
  not "0.9x" or "level 5".

  keep the same node id.

  examples:
    "the left-shape sizes follow a distribute-three cycle of medium, small, and large"
    → subject: "left-shape sizes", predicate: "distribute-three", object: "M, S, L", label: "left size: M-S-L
  cycle"

    "choices 0, 3, and 7 are eliminated because their size does not match"
    → subject: "choices 0,3,7", predicate: "eliminated by", object: "size mismatch", label: "eliminate 0,3,7: size"

    "the answer is choice 5"
    → subject: null, predicate: null, object: null, label: "answer: choice 5"
  \end{Verbatim}


  \paragraph{Output Schema (organized by Zod)}
  \begin{Verbatim}[fontsize=\small]
  {
    "nodes": [
      {
        "id": "...",
        "label": "...",
        "subject": "..." | null,
        "predicate": "..." | null,
        "object": "..." | null
      }
    ]
  }
\end{Verbatim}

\subsubsection{Dataset-Specific Prompt Prefixes}
\label{ap:prompt_refinement}
In the first stage (\ref{ap:system_prompt_QA_reasoning}: question-answer pair to reasoning structure), we prepend dataset-specific prompt prefixes before the general developer prompt. For I-RAVEN, these prefixes describe the matrix layout, valid rule types, configuration-specific structure, and SVG-related constraints.
  For Big-Bench Hard, these prefixes define task-specific reasoning templates for temporal reasoning, logical deduction, and tracking shuffled objects.

\clearpage
\subsection{System Architecture}
\label{ap:system_architecture}
\begin{figure}[H]
\centering
  \includegraphics[width=0.45\textwidth]{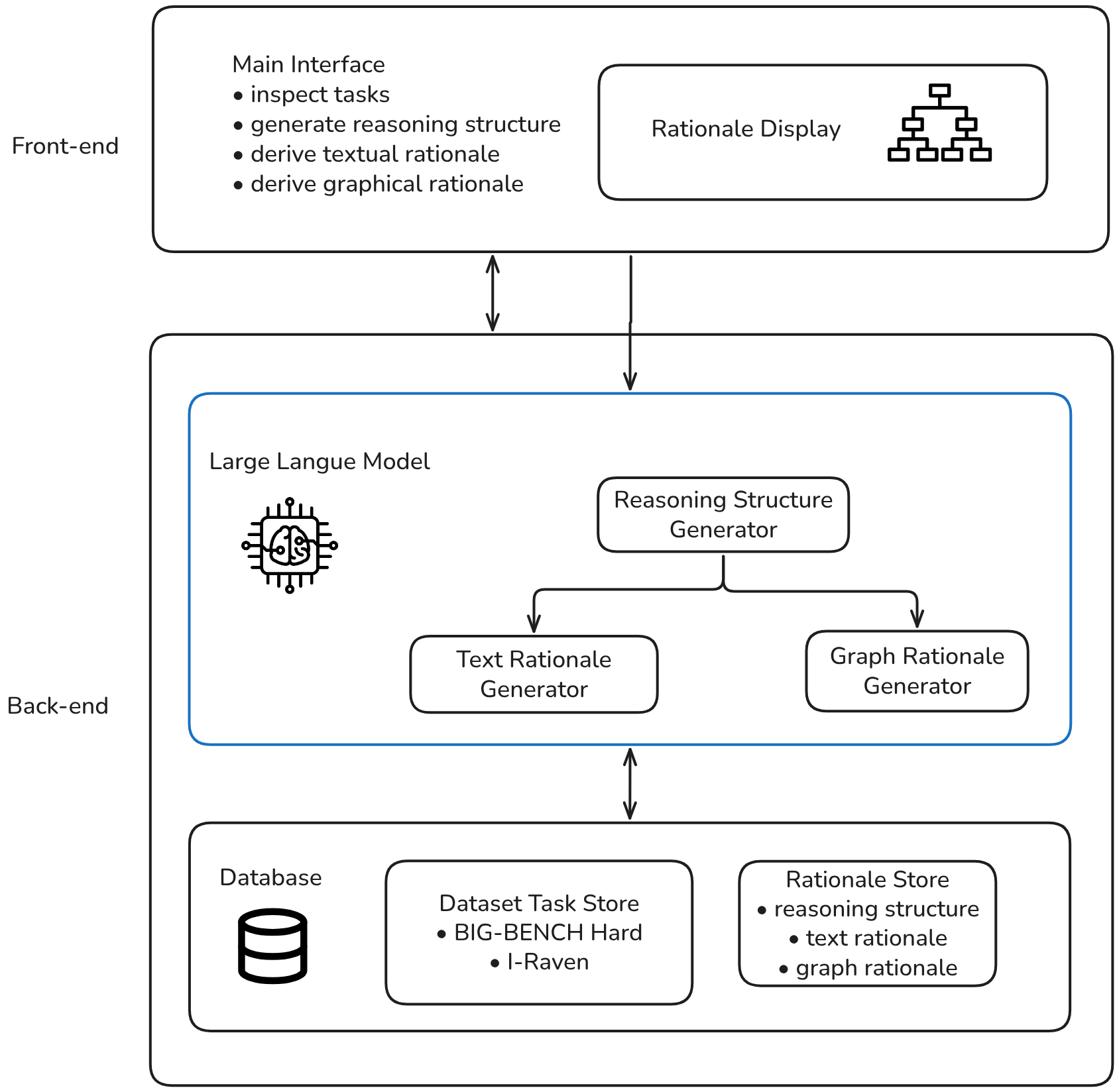}
  \caption{System architecture of \systemName. 
  }
\Description{A system architecture diagram divided into front-end and back-end layers. The front-end contains a Main Interface (supporting task inspection and rationale generation) connected to a Rationale Display panel. The back-end contains an LLM block with three components: a Reasoning Structure Generator at the top, branching into a Text Rationale Generator on the left and a Graph Rationale Generator on the right. Below the LLM block is a Database containing a Dataset Task Store (BIG-Bench Hard and I-RAVEN) and a Rationale Store (reasoning structure, text rationale, graph rationale). Bidirectional arrows connect the front-end to the back-end LLM block, and between the LLM block and the database.}

  \label{fig:sup_system_architecture}
\end{figure}

\subsection{Example of Visually Augmented Verbal Reasoning Rationale}
\label{ap:system_bbh_svg}

\begin{figure}[H]
\centering
  \includegraphics[width=0.95\textwidth]{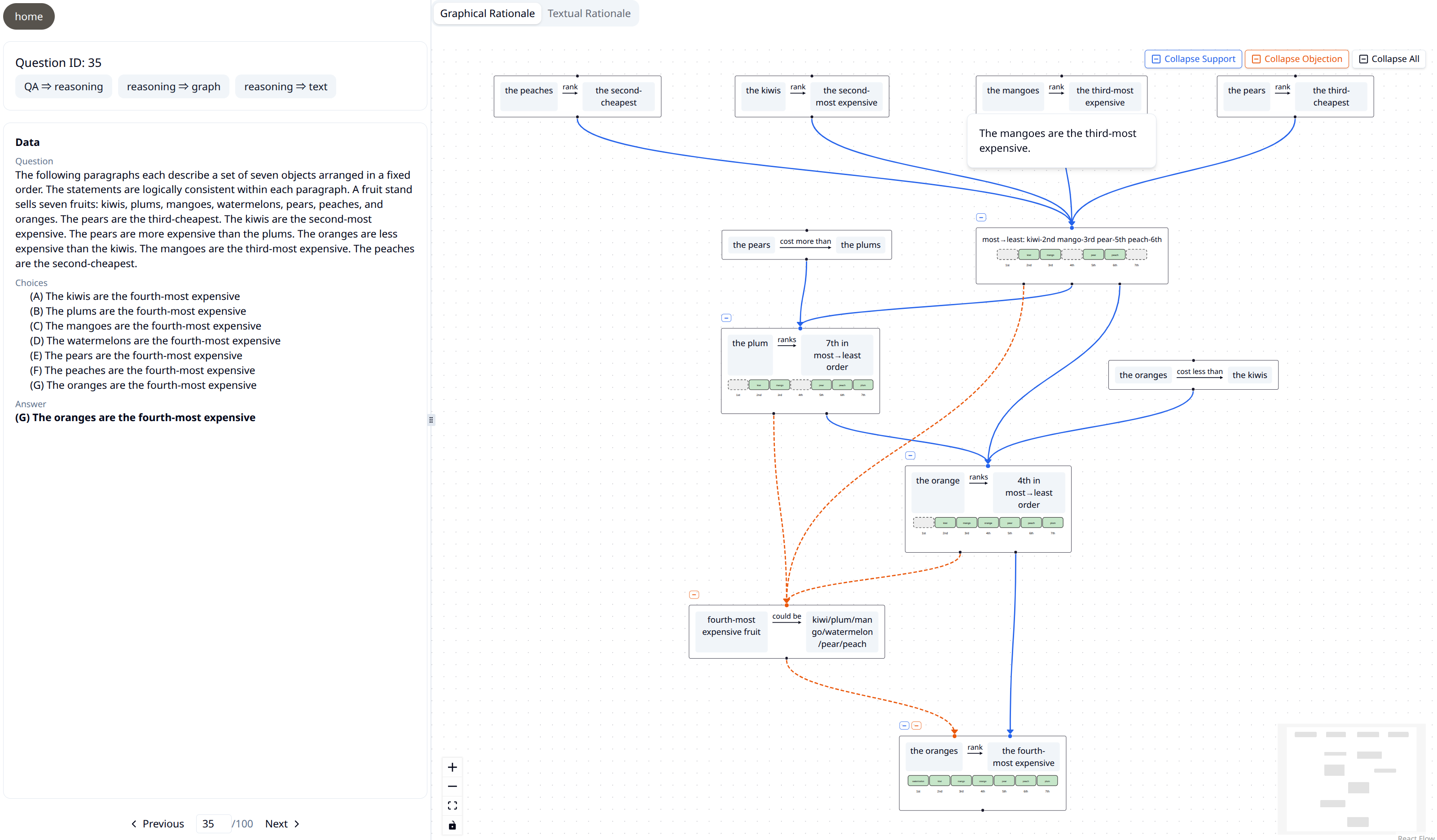}
  \caption{Example of a visually augmented rationale for a BBH Logical Deduction (7 Objects) task. }
  \Description{A screenshot of the \systemName interface demonstrating a visually augmented graphical rationale for a BBH Logical Deduction (7 Objects) task. Unlike the standard graphical rationale that contains text-only nodes, this augmented version embeds small SVG-based spatial diagrams directly inside the argument graph nodes: each oval node displays a miniature linear ordering diagram alongside compressed text, visually encoding the relative positions of fruits (kiwis, plums, mangoes, watermelons, pears, peaches, oranges) on a ranked scale. The left panel shows the task question, eight answer choices (A--G), and the correct answer. The right panel shows the resulting argument graph: a large directed graph whose nodes combine textual reasoning steps with embedded position indicators, connected by blue solid edges (support) and orange dashed edges (objection), converging from multiple premise nodes at the top toward a single final answer node at the bottom.}
\end{figure}

\clearpage
\section{Technical Evaluation Details}
\label{ap:technical_tasks}

\subsection{Example Tasks from BIG-Bench Hard}
\begin{figure}[htbp]
  \centering
  \begin{subfigure}[t]{0.313\linewidth}
    \centering
    \includegraphics[width=\linewidth]{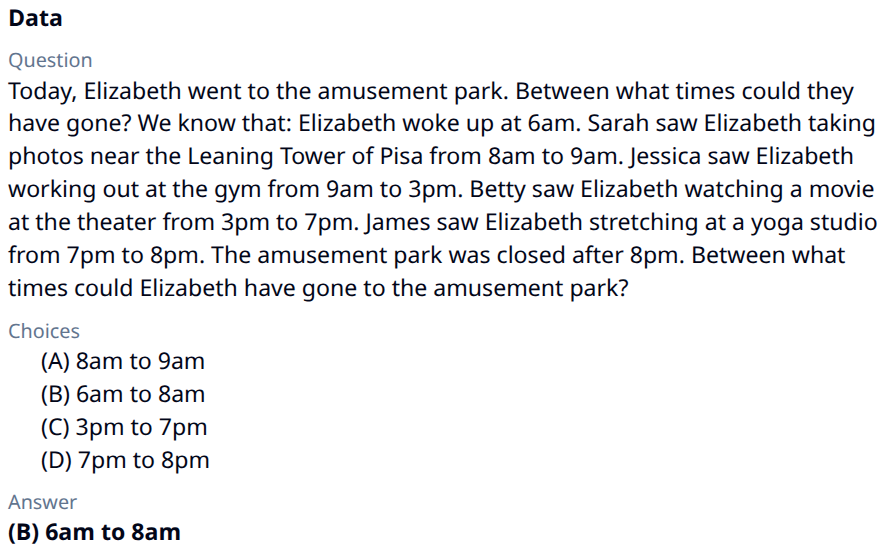}
    \caption{Temporal Sequence}
    \label{fig:bbh-task2}
  \end{subfigure}\hspace{0.005\linewidth}
  \begin{subfigure}[t]{0.313\linewidth}
    \centering
    \includegraphics[width=\linewidth]{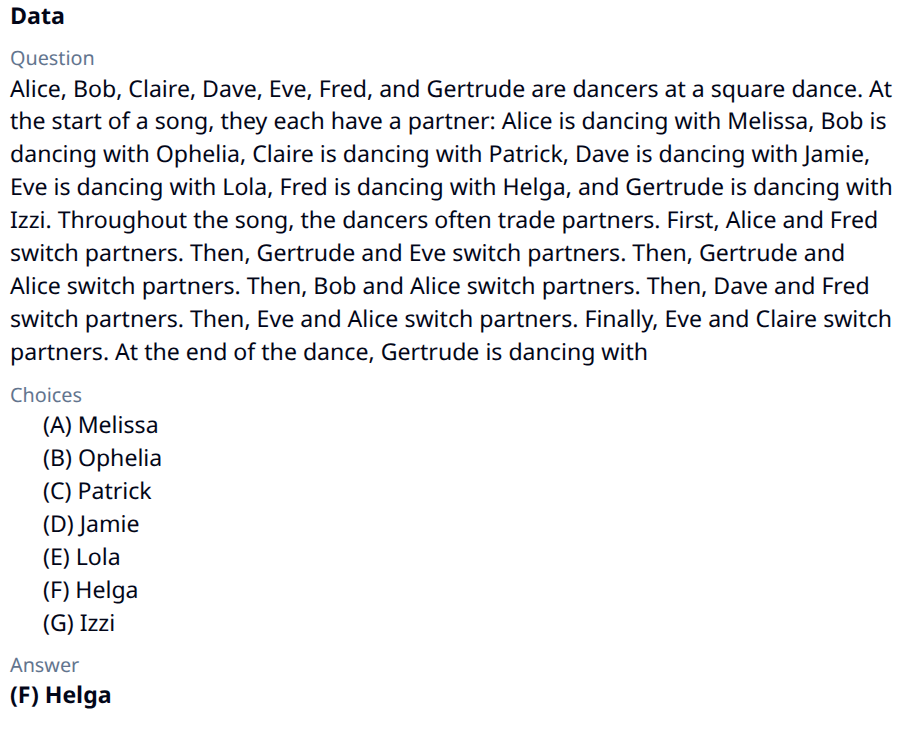}
    \caption{Tracking Shuffled Objects (7
Objects)}
    \label{fig:bbh-task3}
  \end{subfigure}\hspace{0.005\linewidth}
  \begin{subfigure}[t]{0.313\linewidth}
    \centering
    \includegraphics[width=\linewidth]{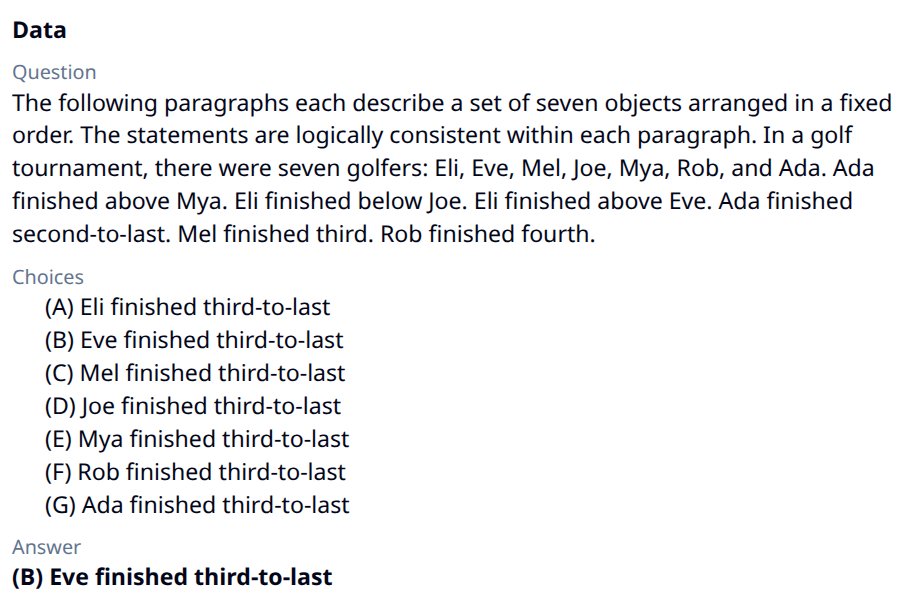}
    \caption{Logical Deduction (7 Objects)}
    \label{fig:bbh-task1}
  \end{subfigure}
  \caption{Three example tasks from Big-Bench-Hard.}
  \Description{Three side-by-side screenshots of example tasks from Big-Bench Hard, each showing a data section, answer choices, and the correct answer. (a) Temporal Sequence: a text passage describing Elizabeth's daily schedule with activities at various times; the correct answer identifies what time a specific activity occurs. (b) Tracking Shuffled Objects (7 Objects): a text passage describing seven dancers at a square dance who repeatedly swap partners; the correct answer identifies who Gertrude is dancing with at the end. (c) Logical Deduction (7 Objects): a text passage stating logical ordering constraints among seven golfers in a golf tournament in a ranked sequence; the correct answer identifies which golfer finished third-to-last. All three tasks present multiple-choice options (four to eight choices) and require multi-step verbal reasoning to solve.}
  \label{fig:bbh-example-tasks}
\end{figure}

\subsection{Example Tasks from I-RAVEN}
\begin{figure}[htbp]
  \centering
  \begin{subfigure}[t]{0.313\linewidth}
    \centering
    \includegraphics[width=\linewidth]{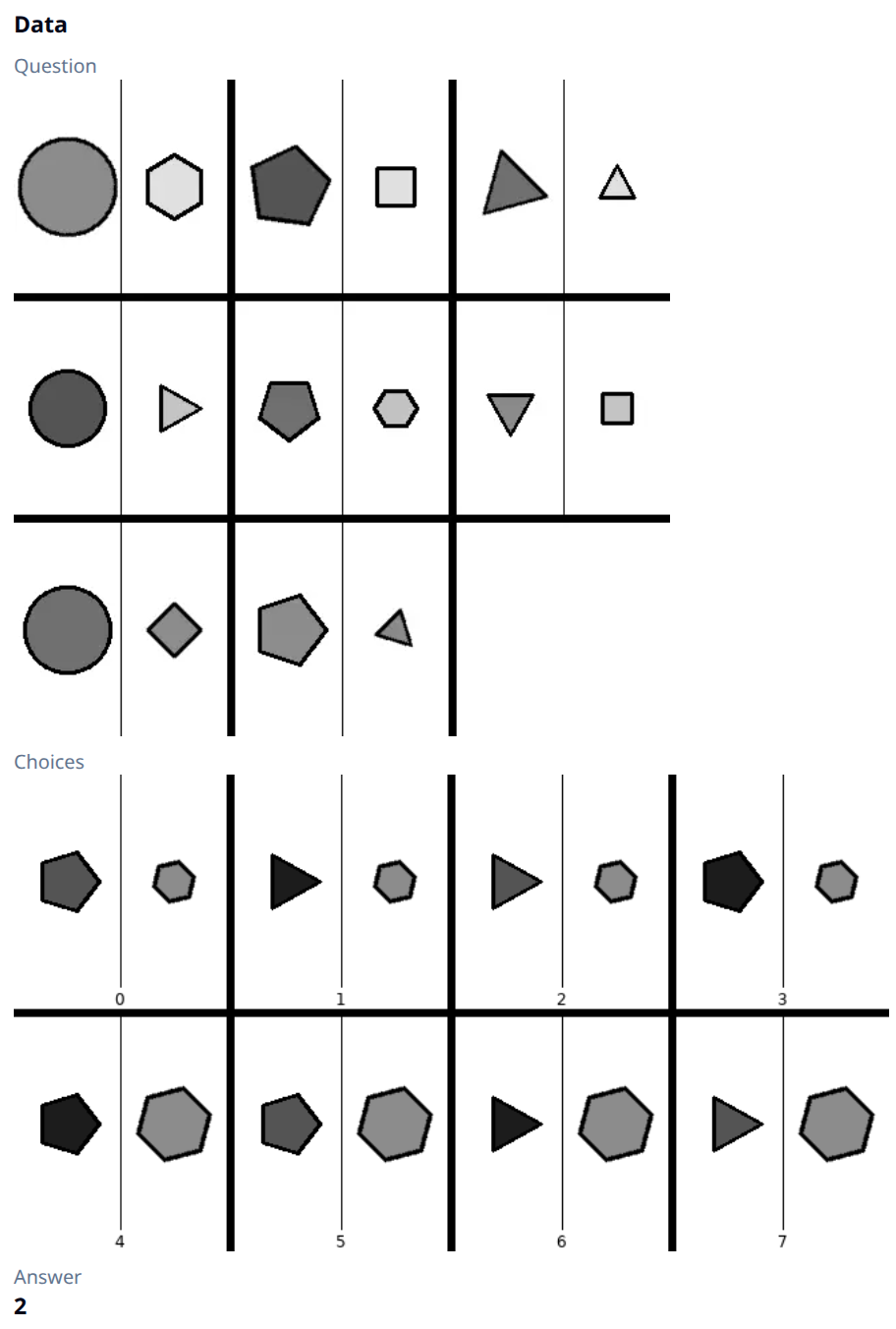}
    \caption{L-R}
    \label{fig:iraven-task3}
  \end{subfigure}
  \hspace{0.005\linewidth}
  \begin{subfigure}[t]{0.313\linewidth}
    \centering
    \includegraphics[width=\linewidth]{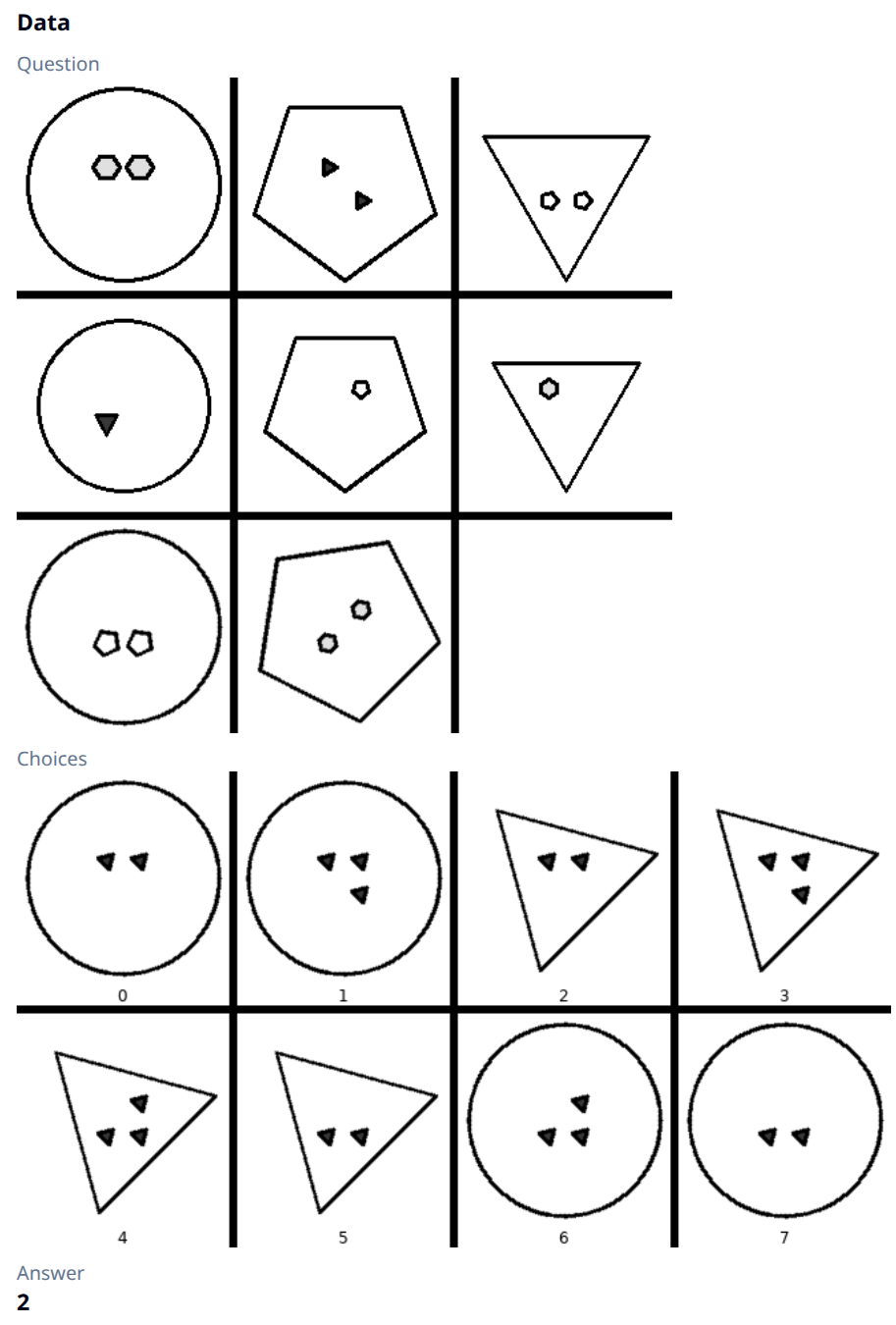}
    \caption{Out-InGrid}
    \label{fig:iraven-task2}
  \end{subfigure}
  \hspace{0.005\linewidth}
  \begin{subfigure}[t]{0.313\linewidth}
    \centering
    \includegraphics[width=\linewidth]{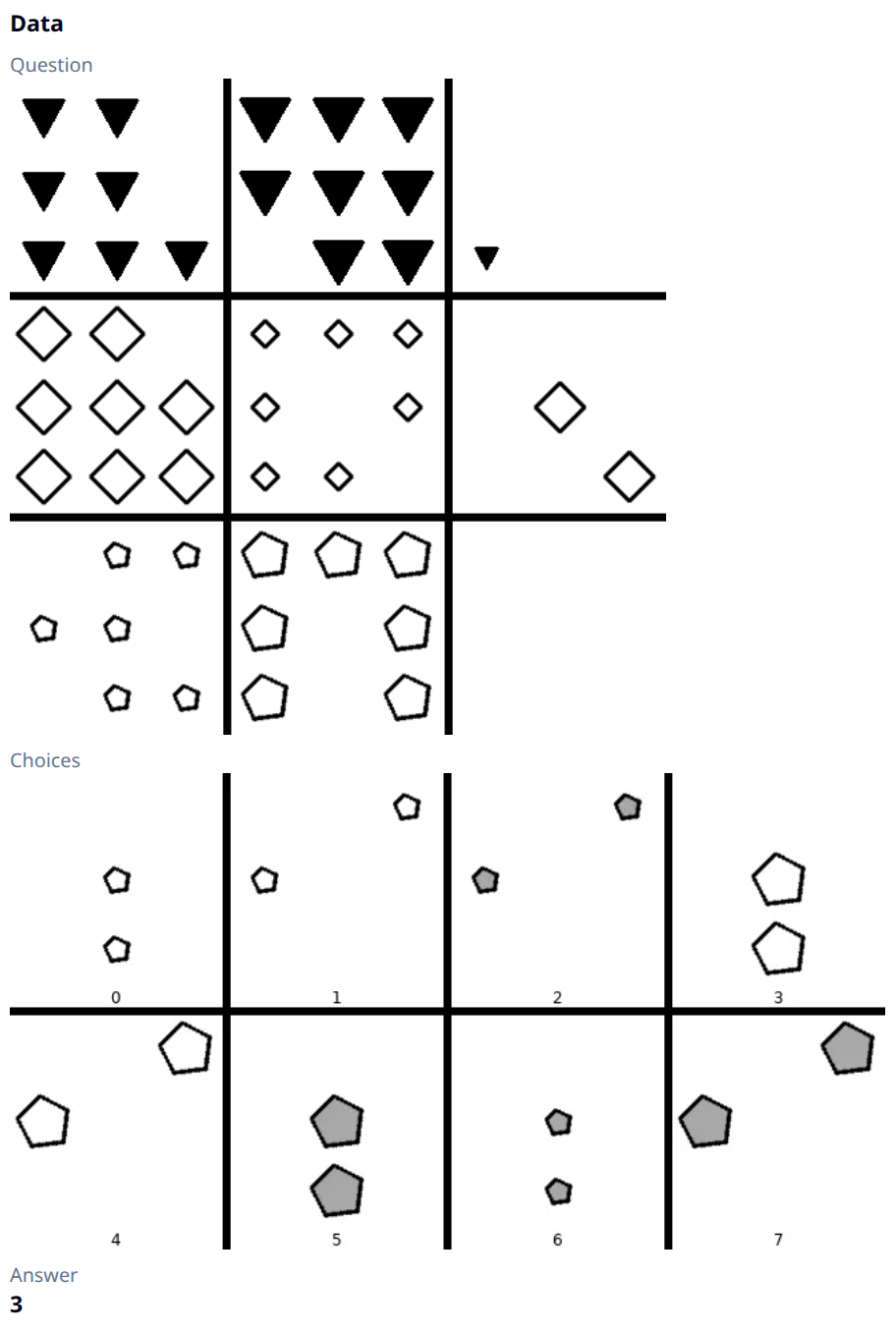}
    \caption{3$\times$3Grid}
    \label{fig:iraven-task1}
  \end{subfigure}
  \caption{Three example tasks from I-RAVEN.}
  \Description{Three side-by-side screenshots of example tasks from I-RAVEN, each showing a 3×3 visual matrix of panels containing geometric shapes, with one panel missing in the bottom row. Participants must select the correct missing panel from eight answer choices shown below the matrix. (a) L-R: each row contains shapes that vary along a left-to-right spatial rule. (b) Out-InGrid: each row contains shapes arranged according to an outside-to-inside grid rule. (c) 3×3Grid: each row contains shapes arranged in a 3×3 sub-grid pattern with varying fills. All three tasks require visual pattern recognition across rows to identify the underlying rule and select the matching missing panel.}
  \label{fig:iraven-example-tasks}
\end{figure}

\clearpage
\section{User Study Details}

\subsection{Prompt for Generating Textual Rationale}
\label{ap:user_prompt_text}
\paragraph{Developer Prompt}
\begin{Verbatim}[fontsize=\small]
you are a reasoning analyst.
given an argument map (JSON with nodes and edges), convert it into a well structured markdown rationale.

**goals**
- the text must read as a natural, self contained explanation, not a list of claims.
- guide the reader through the reasoning step by step so the logic feels intuitive.
- use clear causal connectors (because, therefore, however) to link ideas.

**formatting**
- use markdown headings (##) for the conclusion and key sub-conclusions.
- use bullet points for supporting premises under each heading.
- use "---" dividers between major reasoning sections.
- when a premise is an objection (relation = "objection"), mark it clearly (e.g. "however, ...").
- if a node has an svg field, embed the svg directly in the markdown (not in a code block).
- preserve the logical flow: premises → sub-conclusions → conclusion.
- keep it concise and direct. do not reference node IDs or include JSON/metadata.
\end{Verbatim}

\paragraph{user input}
This stage takes as input the argument map produced in the previous stage.
\begin{Verbatim}[fontsize=\small]
Argument map:
{JSON serialized argument map}
\end{Verbatim}

\subsection{Pre-Study Survey}
\label{ap:user_presurvey}
\begin{figure*}[htbp]
\centering
  \includegraphics[width=0.32\textwidth]{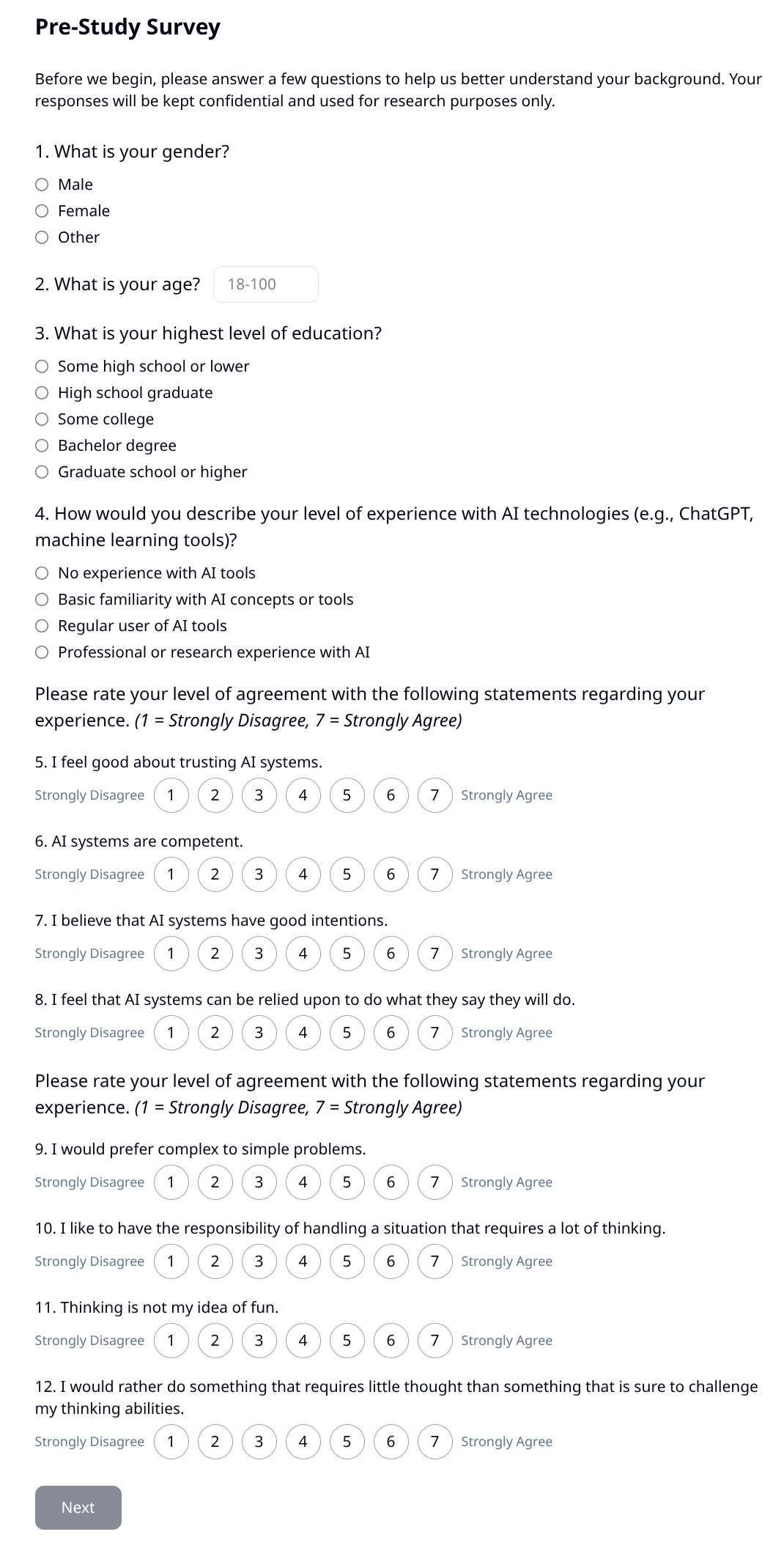}
  \caption{Screenshot of the pre-study survey.}
  \Description{A screenshot of the pre-study survey administered to participants before the main study. The survey contains twelve questions across three sections. Demographic questions (1--4) ask about gender, age, highest level of education, and level of experience with AI technologies such as ChatGPT and machine learning tools. AI attitude questions (5--8) ask participants to rate agreement on a 7-point Likert scale (1 = Strongly Disagree, 7 = Strongly Agree) with statements about trusting AI systems, AI competence, AI good intentions, and AI reliability. Need for Cognition questions (9--12) also use the same 7-point Likert scale.}
  \label{fig:sup_Pre_Study_Survey}
\end{figure*}

\subsection{Post-Study Survey}
\label{ap:user_postsurvey}

\begin{figure*}[htbp]
\centering
  \includegraphics[width=0.34\textwidth]{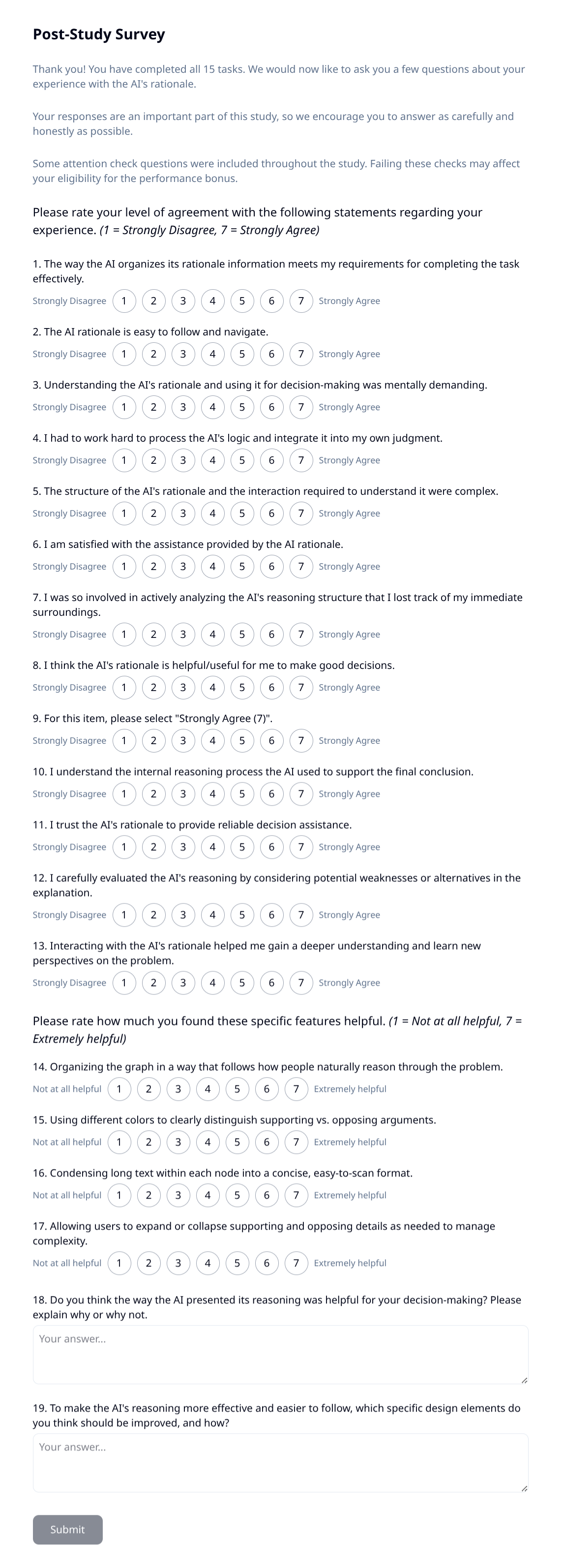}
  \caption{Screenshot of the post-study survey.}
  \Description{A screenshot of the post-study survey administered after participants completed all 15 tasks. The survey contains 19 items across three sections. Subjective experience questions (1--13) use a 7-point Likert scale (1 = Strongly Disagree, 7 = Strongly Agree) and cover: Usability, Task Load, Satisfaction, Engagement, Helpfulness, an attention check item instructing participants to select ``Strongly Agree,'' Understanding, Trust, Critical Thinking, and Learning Gain. Feature helpfulness questions (14--17) use a 7-point scale (1 = Not at all helpful, 7 = Extremely helpful) and ask about four specific design features: Alignment with Human Decision Logic, Semantic Color-Coding, Text Compression, Progressive Disclosure. Open-ended questions (18--19) ask participants to explain whether the AI's reasoning presentation was helpful for decision-making and to suggest specific design improvements.}
  \label{fig:sup_Post_Study_Survey}
\end{figure*}


\end{document}